\documentclass[useAMS,usenatbib]{mn2e}

\usepackage{graphics,graphicx}
\usepackage{times}
\usepackage{amssymb,amsmath}

\newcommand{\kms}{\hbox{km\,s$^{-1}$}}
\newcommand{\Hi}{\hbox{H\,{\sc i}}}

\newcommand{\Ha}{\hbox{H\,$\alpha$}}
\newcommand{\Htwo}{\hbox{H$_{\rm 2}$}}

\newcommand{\Rthtw}{\hbox{$R_{\rm 3-2/2-1}$}}
\newcommand{\Rthon}{\hbox{$R_{\rm 3-2/1-0}$}}
\newcommand{\Rtwon}{\hbox{$R_{\rm 2-1/1-0}$}}
\newcommand{\mf}{\hbox{M\,51}}
\newcommand{\pah}{\hbox{PAH~8\micron}}
\newcommand{\kkms}{\hbox{K\,km\,s$^{-1}$}}

\newcommand{\ta}{\hbox{$T_{\rm A}^{*}$}}
\newcommand{\tmb}{\hbox{$T_{\rm mb}$}}
\newcommand{\ico}{\hbox{$I_{\rm CO}$}}

\newcommand{\xco}{\hbox{$X_{\rm CO}$}}

\newcommand{\xcothree}{\hbox{$X_{\rm CO(3-2)}$}}
\newcommand{\cothree}{\hbox{CO $J$=3-2}}
\newcommand{\cotwo}{\hbox{CO $J$=2-1}}
\newcommand{\coone}{\hbox{CO $J$=1-0}}
\newcommand{\jthree}{\hbox{$J$=3-2}}
\newcommand{\jtwo}{\hbox{$J$=2-1}}
\newcommand{\jone}{\hbox{$J$=1-0}}

\title{A \cothree\/ map of \mf\/ with HARP-B: Radial properties of the spiral structure}
\author[C. Vlahakis et al.]{C. Vlahakis$^{1,2,3}$\thanks{E-mail:
cvlahaki@alma.cl (CV)},  P. van der Werf$^{3}$, F. P. Israel$^{3}$ and R. P. J. Tilanus$^{3,4}$\\
$^{1}$Joint ALMA Observatory / European Southern Observatory, Alonso de Cordova 3107, Vitacura, Santiago, Chile\\
$^{2}$Departamento de Astronomia, Universidad de Chile, Casilla 36-D, Santiago, Chile\\
$^{3}$Leiden Observatory, Leiden University, P.O. Box 9513, 2300 RA Leiden, Netherlands\\
$^{4}$Netherlands Organisation for Scientific Research, P.O. Box 93138, 2509 AC, The Hague, Netherlands
}
\begin{document}

\date{}

\pagerange{\pageref{firstpage}--\pageref{lastpage}} \pubyear{2013}

\maketitle

\label{firstpage}

\begin{abstract}
We present the first complete \cothree\/ map of the nearby grand-design spiral galaxy \mf\/ (NGC~5194), 
at a spatial resolution of $\sim$600\,pc, obtained with the HARP-B instrument 
on the James Clerk Maxwell Telescope. The map covers the entire optical galaxy disk and out to the 
companion NGC~5195, with \cothree\/ emission detected over an area of  $\sim$9\arcmin$\times$6\arcmin 
($\sim$21$\times$14\,kpc).
We describe the \cothree\/ integrated intensity map and combine our results with maps of 
\cotwo, \coone\/ and other data from the literature to investigate the variation of the 
molecular gas, atomic gas and polycyclic aromatic hydrocarbon (PAH) properties of 
\mf\/ as a function of distance along the spiral structure on sub-kiloparsec scales.
We find that for the \cothree\ and \cotwo\/ transitions there is a clear difference between 
the variation of arm and inter-arm emission 
with galactocentric radius, with the inter-arm emission relatively constant with radius 
and the contrast between arm and inter-arm emission decreasing with radius.
For \coone\/ line and \Hi\/ emission the variation with radius shows a 
similar trend for the arm and inter-arm regions, and the arm--inter-arm contrast appears 
relatively constant with radius.
We investigate the variation of CO line ratios ($J$=3-2/2-1, $J$=2-1/1-0 and 
$J$=3-2/1-0) as a function of distance along the spiral structure.  
Line ratios are consistent with the range of typical values for other nearby galaxies 
in the literature. 
The highest CO~\jthree/\jtwo\/ line ratios are found in the central $\sim$1\,kiloparsec 
and in the spiral arms and the lowest line ratios in the inter-arm regions.
We find no clear evidence of a trend with radius for the spiral arms but for the 
inter-arm regions there appears to be a trend for all CO line ratios to increase with radius.
We find a strong relationship between the ratio of 
\cothree\/ intensity to stellar continuum-subtracted 8\micron\/ PAH surface brightness and the 
\cothree\/ intensity that appears to vary with radius. 
\end{abstract}


\section{Introduction}
\label{sec:intro}

M\,51 (NGC~5194), the `Whirlpool galaxy', is a nearby grand-design spiral galaxy 
viewed almost face-on (Table~\ref{properties}), 
with two prominent spiral arms that can be traced close 
to the nuclear region, and is undergoing an interaction with a small companion, 
NGC~5195, 4.5\arcmin\/ to the north-east. With its large angular size and strong spiral 
structure \mf\/ is well studied across a broad 
range of wavebands from X-ray to radio, 
and has been the subject of numerous studies investigating the origin 
of grand design spiral structure \citep[e.g.][]{tully74,garcia-burilloetal93b,shetty07,meidtetal08} 
and the relationship between spiral arms and star formation 
\citep[e.g.][]{hitschfeld09,foyleetal10}.
It is a Seyfert 2 \citep{hoetal97} and hosts a low-luminosity active galactic 
nucleus (AGN) surrounded by a $\sim$100\,pc disk of warm dense gas 
\citep{kohnoetal96,matsushitaetal98}. The two prominent spiral arms in \mf, presumed to 
be caused via tidal interaction with the companion, are rich in molecular gas 
\citep[e.g.][]{aalto99,shetty07,koda09}, with molecular gas dominant over atomic gas until 
the outskirts of the disk \citep[e.g.][]{garcia-burilloetal93a,nakaietal94,schuster07}.

Molecular clouds provide the fuel for star formation, and in nearby galactic disks 
knowledge of the distribution of molecular hydrogen gas is a crucial tool for 
investigating star formation and spiral structure. Since molecular hydrogen 
is difficult to detect directly, as it lacks a permanent dipole 
moment, carbon monoxide (CO), observable at (sub)millimetre (submm) wavelengths, is widely used as a 
tracer of molecular hydrogen gas both in molecular clouds in the Milky Way and in 
external galaxies. At a distance of 8.2\,Mpc \citep[][1\arcsec\/ corresponds to 
$\sim 40$~pc]{tully74} and an inclination $i = 20\degr$ \citep{tully74}, the proximity and 
face-on nature of \mf\/ makes it possible to use single-dish (sub)mm telescopes 
to study spiral arm structures and sub-structures at spatial resolutions of a few 
hundred parsecs. At this spatial resolution we are not sensitive to individual molecular 
clouds but rather clumps of molecular clouds, or Giant Molecular Associations (GMAs), 
as well as diffuse gas. The spiral arms of \mf\/ contain numerous GMAs 
\citep[as many as 16 are noted by][]{aalto99}

Since stars form not from the 
diffuse envelopes of molecular clouds but from their dense 
molecular cores, tracing the warm dense gas is important for our understanding 
of star formation. The $J$=3-2 transition of CO (the $J$=3 level is 33\,K above ground state) 
is a tracer of this warm, dense gas, while the lower critical density \coone\/ transition 
(the $J$=1 level is 5.5 K above ground state) is sensitive to the majority of gas in the 
cold ISM, including, for example, more diffuse gas. In fact, many authors have shown that 
the \cothree\/ and \coone\/ lines cannot be tracing exactly the same components of 
molecular gas, and, importantly, that \cothree\/ is the better tracer of warm dense gas 
(and hence star forming potential) \citep[e.g.][and references therein]{wilson09}.

Although \mf\/ has been mapped in low-$J$ rotational $^{12}$CO and $^{13}$CO transitions in a number 
of single-dish and interferometric studies, until now complete mapped observations of the 
full disk of \mf\/ in the higher energy level transition \cothree\/ line have been 
lacking, in particular because of the need for good telescope sites for higher frequency 
observations. Single-dish maps of the \coone\/ and \cotwo\/ transitions were presented by 
\citet[][]{garcia-burilloetal93a}, \citet[][]{garcia-burilloetal93b}, 
\citet[][]{schuster07},  \citet{nakaietal94} and \citet[][]{koda09,koda11}  (using the IRAM 30m and Nobeyama 
telescopes), and aperture synthesis maps have been presented by e.g. \citet{loetal87}, 
\citet{sakamotoetal99}, \citet{aalto99}, \citet{reganetal01} and \citet[][]{koda09,koda11}. 
These studies have indicated that CO is mainly confined to the spiral arms. 
Other authors have studied tracers such as CO, CI, and CII in the central region 
or at individual positions in the spiral arms 
\citep[e.g.][]{israelandbaas02,kramer05,israeletal06}. \cothree\/ observations to date 
have been restricted to the central regions and inner spiral arms of the galaxy 
\citep[e.g.][]{wielebinski99,dumkeetal01} and have also had a relatively coarse angular 
resolution. Aperture synthesis mapping of the central region has also been carried out 
\citep[e.g.][]{matsushitaetal04,schinnerer10}, but aperture synthesis maps 
are known to resolve out and miss both a 
significant proportion of the CO flux and any extended emission. Single-dish \cothree\/ 
observations of the full disk of \mf\/ are thus essential for determining the true 
distribution of molecular gas including any diffuse emission in the inter-arm regions, 
and thus its relation to star formation and the total reservoir of star-forming gas. 

As well as understanding the distribution of molecular gas as a function of the spiral structure, 
it is also important to understand the relationship across multiple wavebands, from observations of 
the stellar component through to the dust and atomic gas components of the interstellar medium (ISM).
Previous studies have suggested that polycyclic aromatic hydrocarbon (PAH) emission in the mid-infrared 
is a tracer of dust and gas in the cool ISM. 
For example, \citet[][]{reganetal06} found that within nearby galaxies PAH emission is correlated with 
CO line emission, and several studies using {\it Spitzer} data have found that PAH emission is 
correlated with cool ($\sim$20~K) dust emission in nearby spiral galaxies \citep[e.g.][]{bendo06,zhuetal08}. 
However, using HARP \cothree\/ observations of the spiral galaxy NGC~2403, \citet[][]{bendo10} found the scale 
length of the \cothree\/ radial profile to be statistically identical to the scale length of the 
continuum-subtracted 8\micron\/ (PAH) emission but found no correlation on sub-kpc scales.

In this paper we present the first complete \cothree\/ map of \mf\/ 
covering the entire optical galaxy disk over an area of $\sim$10\arcmin$\times$7\arcmin\/ 
($\sim$24$\times$17\,kpc) at a spatial resolution of $\sim$600\,pc.
The paper is organised as follows. 
In Section~\ref{sec:obs-data} we describe our HARP-B \cothree\/ observations, the 
data reduction and processing of the HARP-B data, and our treatment of 
ancillary data we took from the literature.
Sections~\ref{sec:images}--\ref{sec:spiral} present our results. 
In Section~\ref{sec:images} we present the HARP-B \cothree\/ map, describe the 
distribution of the \cothree\/ emission in a 
qualitative way, and compare our \cothree\/ map 
to maps of other tracers of molecular gas (\cotwo, \coone), atomic gas (\Hi), star formation 
(\Ha, 24\micron) and PAH 8\micron\/ emission using data available in the literature. 
We also investigate the contamination of submm continuum emission by in-band \cothree\/ line emission. 
In Section~\ref{sec:geometry} we discuss how the appearance of \mf's bright molecular gas 
rich spiral arms varies for different projected orientations and
in Section~\ref{sec:coratio} we disucss maps of the CO line ratio.
In Section~\ref{sec:spiral} we examine and discuss the distribution and properties 
of the \cothree\/ emission 
as a function of the distance along the spiral arms and inter-arm regions, and compare this 
to the distribution seen for other wave bands.

\section{Observations and data reduction}
\label{sec:obs-data}

\begin{table*}
\centering
\begin{minipage}{140mm}
\caption{Properties and observing parameters of \mf}
\begin{tabular}{ll}
\hline
Parameter & Value\\
\hline
Centre R. A. (J2000) & 13$^{h}$29$^{m}$52.7$^{s}$\\
Centre Decl. (J2000) & $+$47$\degr$11\arcmin43\arcsec\\
Morphological type$^{a}$ & SA(s)bc pec\\
Distance $^{b}$ & 8.2~Mpc \\
Position angle$^{b}$& 170$\degr$ \\
Inclination$^{b}$ & 20$\degr$ \\
$V_{LSR}$$^{b}$ & $+$464\,\kms \\
\smallskip
Bandwidth & 1~GHz\\
Original channel width & 0.488 MHz (0.423~\kms)\\
rms (\ta) per 0.423\,\kms\/ channel &  0.037 K \\
rms (\ta) per 10.6\,\kms\/ velocity-binned channel &  0.008 K \\
Beamsize & 15\arcsec \\
Spatial scale  & 600 pc/beam \\ 
\hline
\end{tabular}\\
$^{a}$ \citet[][]{devaucouleurs91}\\
$^{b}$ \citet[e.g.][]{tully74}
\label{properties}
\end{minipage}
\end{table*}

\subsection{HARP-B CO J=3-2 data} 
\label{sec:co32-data}
Observations of the J=3-2 transition of $^{12}$CO (rest frequency 345.79 GHz) were carried out 
with the 16-pixel array receiver HARP-B \citep[][]{buckle09} at the James Clerk Maxwell 
Telescope (JCMT) in January 2007.
The telescope beam size (Half Power Beam Width) at this frequency is 
$\sim$15\arcsec, which corresponds to $\sim$600\,pc at a distance of 8.2\,Mpc.
 The 16 receivers of the HARP-B array are separated by 
approximately two beam widths ($\sim$30\arcsec), and a ``basket-weave'' scan pattern 
was used to achieve a final map sampling with spacing of half a beam-size. 
HARP-B was used in conjunction with the Auto-Correlation Spectral Imaging System (ACSIS) backend spectrometer, 
which was configured to have a bandwidth of 1~GHz and a resolution of 0.488 MHz (0.423 \kms at the observed 
frequency) 
per channel.

The data reduction was carried out using 
Starlink\footnote{The Starlink software package \citep{currie08} is available from http://starlink.jach.hawaii.edu/} software packages such as 
KAPPA, SMURF and CUPID in a similar manner to that described in detail in 
\citet{wilson09} and \citet{warren10}. Here, we give a brief overview of the data reduction process. 
First, the individual raw data files were inspected; obviously noisy channels 
were removed, the data was despiked, and any bad baselines were flagged.
Second, the raw scans were combined into a data cube using the task {\it makecube} that 
is part of the SMURF software package. In this process, a sinc($\pi~x$)sinc($k\pi~x$) kernel 
was used to determine the contribution of of individual receiver pixels to each map pixel 
(whose size was chosen to be 7.5\arcsec); this ``SincSinc'' function is an often used weighting 
function that results in a significant reduction in the noise in the map while having only a minor effect on the resolution. A baseline was then removed by fitting a fifth-order 
polynomial to line-free regions 
of the data cube. Thirdly, we created a mask of regions identified as having emission with 
a signal-to-noise $>$3 in a boxcar smoothed data cube (smoothed by 3$\times$3 pixels 
spatially and over 25 velocity channels), using the task {\it findclumps} that is part 
of the CUPID software 
package and uses the {\it clumpfind} algorithm \citep[][]{williams94}. 
Lastly, we applied this mask to the original high resolution data cube and from this 
produced moment maps. The  CO integrated intensity (zeroth order moment) and 
velocity field (first order moment ) maps are shown in Figure~\ref{fig:maps-ico32}. 
It is the \cothree\/ integrated intensity map that is the focus of this paper.

The size of the \cothree\/ map is 
$\sim$10\arcmin$\times$7\arcmin\/ (or $\sim$24~kpc$\times$17~kpc), 
which covers the full galaxy disk. \cothree\/ emission is detected over the large proportion of 
this area (see Section~\ref{sec:images}). The map rms noise level, measured over the main galaxy disk, 
is $\sim$37\,mK (on the corrected antenna temperature, \ta, scale) 
for a channel with 0.423\,\kms\/ velocity resolution (or $\sim$8\,mK for a 10.6\,\kms\/ channel).  
We note that pixels at the very edges of the map tend to exhibit an increased noise level, and we thus 
exclude these from our analysis. 
The 1$\sigma$ noise level in the total intensity map is estimated to be $\sim$0.1~K\,\kms 
(in \ta\/ units),
based on the rms noise per 0.423\,\kms\/
channel and the number of channels at a typical position of emission.
To convert the data from \ta\/ units to main beam brightness 
temperature (\tmb) units we divided by the main beam efficiency $\eta_{mb}$=0.6. The 
uncertainty on  $\eta_{mb}$ is 10--15\,\%. 
Throughout the paper, all results will 
be quoted in \tmb\/ units.

We compared our \cothree\/ map to previous \cothree\/ Heinrich-Hertz Telescope (HHT) data: 
a map of the inner 3\arcmin--4\arcmin\/ presented by \citet{wielebinski99} and a single HHT 
pointing at the central position presented by \citet{mauersberger99}. 
After smoothing our map slightly to account for the 
difference in beam-size (the HHT beam-size was 22\arcsec and 18\arcsec\/ for the former and 
latter observations, respectively), we confirmed that both the shape of the spectra and the 
integrated intensity values in our map are in agreement with these previous studies within the 
calibration uncertainties (the uncertainty on $\eta_{mb}$ is 20\% 
for the HHT data). We note that contrary to the findings of \citet{wielebinski99}, our HARP-B data 
does not show evidence of particularly extended \cothree\/ emission, and it appears that the HHT data 
may to some extent have been smeared out by an error beam.

We also compared the integrated intensity values in the \cothree\/ HARP-B map to previous JCMT
\cothree\/ observations of the central regions presented by \citet{israeletal06} and find very good agreement.

\subsection{Ancillary data}
\label{sec:ancillary}

\subsubsection{\cotwo\/ and \coone\/ data}
\label{sec:ancillary:co}

For \cotwo\/ and \coone, we used the integrated intensity maps presented in \citet{schuster07} and \citet[][]{koda09,koda11}, 
respectively, which were kindly provided by the authors. 

The \cotwo\/ dataset was observed with the HERA instrument on the IRAM 30-m telescope, 
whose beamsize at the frequency of the \cotwo\/ line is $\sim$12\arcsec. 
The mean baseline rms is 18\,mK (\ta\/ scale) at 5\,\kms\/ velocity resolution and the 
1$\sigma$ noise level of the \cotwo\/ map is 0.65\,K\,\kms, on the \ta\/ scale. To convert 
from \ta\/ units to main beam units we divide by $\eta_{mb}$=0.57 \citep[][]{schuster07}.  
For further details of the \cotwo\/ observations we refer to \citet[][]{schuster07}.

The \coone\/ dataset was observed with the BEARS instrument on the Nobeyama Radio Observatory 45-m telescope, 
whose beamsize at 115\,GHz is 15\arcsec. The 1$\sigma$ noise level is 14.7\,mK (\ta\/ scale) in a 10\,\kms\/ channel. 
To convert from \ta\/ units to main beam units the data is divided by $\eta_{mb}$=0.4. For further details of the 
\coone\/ data we refer to \citet[][]{koda09,koda11}.

As for the \cothree\/ map, we also compared the \cotwo\/ and \coone\/ maps to previous datasets 
and again find good agreement.

\subsubsection{\Hi\/ data}
\label{sec:ancillary:hi}
We took \Hi\/ integrated intensity maps from The \Hi\/ Nearby Galaxy Survey \citep[THINGS;][]{walter08}, 
via the THINGS website\footnote{http://www.mpia-hd.mpg.de/THINGS/Overview.html}. 
THINGS was observed with the Very Large Array (VLA) in multiple configurations to provide uniform 
sensitivity over a range of spatial scales, and provides two differently weighted datasets -- a natural 
weighted integrated intensity map and 
a robust weighted map. In this work we use the natural weighted map as this should be more sensitive to 
extended, large-scale, emission and we do not require the higher resolution provided by the robust weighted map. 
The Full Width at Half Maximum (FWHM) of the Point Spread Function (PSF) of the \Hi\/ image is 11.92\arcsec$\times$10.01\arcsec.

\subsubsection{\Ha\/ and {\it Spitzer Space Telescope} data}
\label{sec:ancillary:sings}
We used \Ha\ (R-band subtracted), {\it Spitzer} Infrared Array Camera (IRAC) 3.6 and 8\micron\/ and 
{\it Spitzer} Multiband Imaging Photometer (MIPS) 24\micron\/ and 70\micron\/ data from the {\it Spitzer} 
Infrared Nearby Galaxies Survey 
\citep[SINGS;][]{kennicutt03} sample, obtained via the 
NASA/IPAC Infrared Science Archive\footnote{http://irsa.ipac.caltech.edu/data/SPITZER/SINGS/}. 
For a description of the data reduction technique we refer to \citet[][]{reganetal06}.
The FWHM of the PSFs are 1.7\arcsec, 2\arcsec, 6\arcsec\/ and 18\arcsec\/ for the 3.6, 8, 24 and 
70\micron\/ data, respectively. 

We used the 3.6 and 8\micron\/ images to produce a map of the PAH 8\micron\/ emission using 
the relationship
\begin{equation}
I_\nu(PAH~8\micron)=I_\nu (8\micron)-0.232 I_{\nu}(3.6\micron),
\end{equation}
derived by \citet[][]{helou04}, to subtract the stellar continuum from the 8\micron\/ image 
(where $I_{\nu}$ has units of MJy\,sr$^{-1}$). For the two IRAC images used in this equation 
we first subtracted residual backgrounds, by interpolating a smoothed version of the background 
outside the optical disk, masked out bright foreground stars, and corrected 
to an `infinite aperture' using values of 0.944 and 0.737, respectively, as described in 
\citet[][]{reachetal05}.

For the 24\micron\/ image we also subtracted a residual background, using the same method 
of interpolating a smoothed version of the background outside the optical disk as we used 
for the IRAC images.

\begin{figure*}
\centering
\includegraphics[width=17cm, clip]{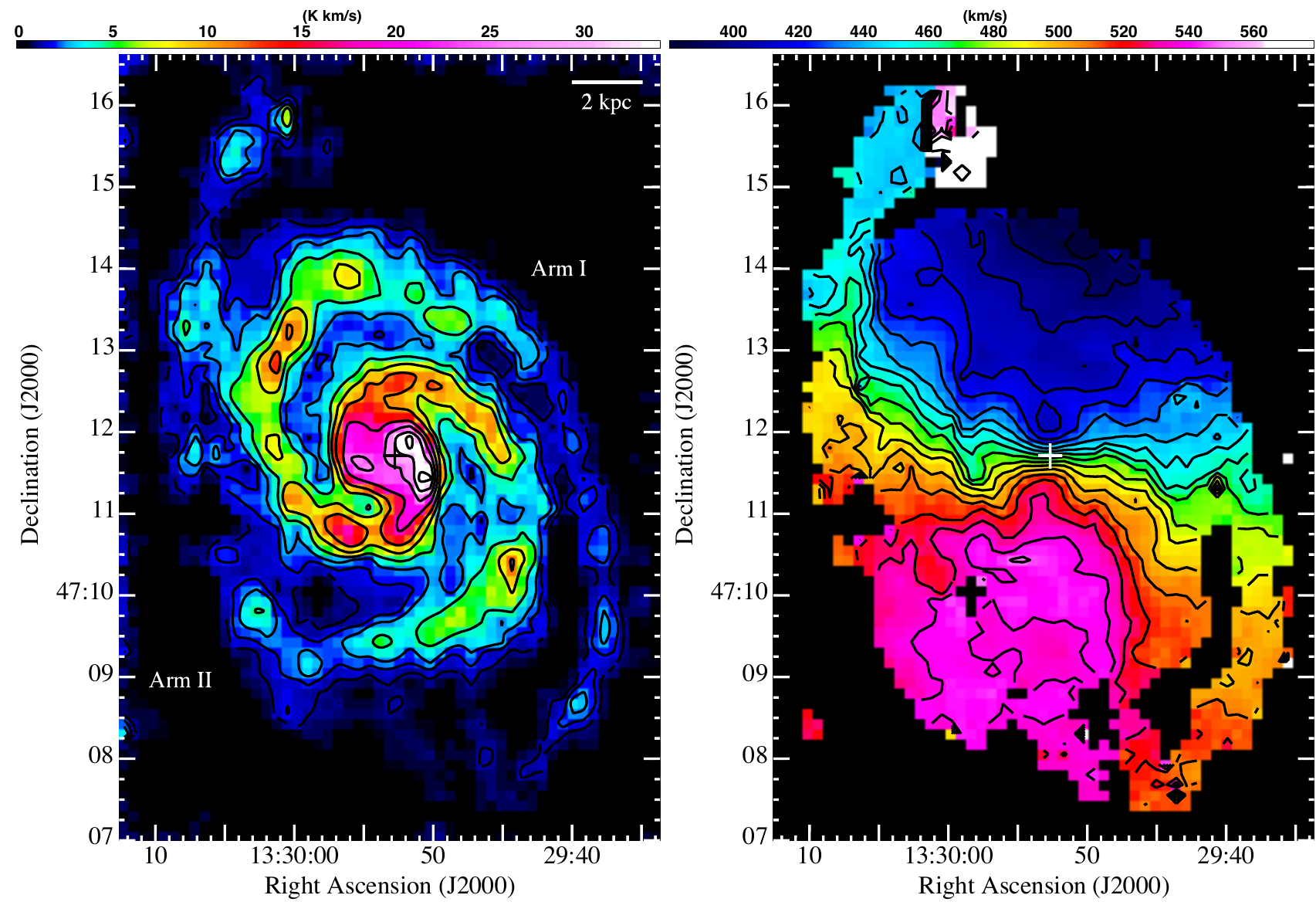}
\caption{Left: HARP-B \cothree\/ integrated intensity (zeroth moment) map of M~51, on the \tmb\/ 
scale (K\,\kms), shown on a logarithmic colour scale. 
Pixels where no  emission was detected are shown in black.
The field of view (FOV) is $\sim$10\arcmin$\times$7\arcmin\/ (or $\sim$24~kpc$\times$17~kpc).
The central position, listed in Table~\ref{properties}, is marked with a cross.
Contours are at 0.5, 1, 2, 4, 7, 11, 18, 25, 32 and 39~\,K\,\kms; the 1$\sigma$ noise level is 
0.17\,K\,\kms). 
The peak integrated intensity is 41.3\,K\,\kms. The beam-size is $\sim$15\arcsec\/ and pixels are 
7.5\arcsec\/ in size (which corresponds to $\sim$300\,pc at the distance of \mf).
Right: HARP-B \cothree\/ velocity field (first moment) map. Contour levels range from 380 to 580~\kms\/ in 
intervals of 10~\kms.
}
\label{fig:maps-ico32}
\end{figure*}

\subsection{Alignment, convolution and re-gridding}
\label{sec:images:alignment}

Since in this work we wish to make comparisons between physical regions 
probed at different wavelengths, for example comparing structures in our \cothree\/ image 
with \cotwo\/ and \coone\/ images taken from the 
literature (see Section~\ref{sec:ancillary:co}), it is important to ensure the image coordinates are aligned and 
to convolve all the images to the lowest common resolution. 
We checked the relative pointing offsets
between the three datasets by comparing peaks in the integrated intensity maps convolved to the 
HARP-B beamsize and pixel grid, by comparing the three datasets to the higher resolution
BIMA SONG \coone\/ map \citep[][]{helfer03}, and by examining the pixel--pixel relationship of 
the integrated intensity maps resampled onto a smaller pixel grid.
We found no obvious offset between the two \coone\/ images, 
and also no obvious offset between the \coone\/ 
images and the \Ha\/ and 24\micron\/ images.
For the \cothree\/ and \cotwo\/ images we found offsets of $+4$\arcsec\/ and $-4$\arcsec, respectively, 
which is consistent with the pointing uncertainties for each dataset. A similar offset for the \cotwo\/ dataset was 
previously reported by \citet{koda12}. We applied these shifts to the HARP and 
HERA data accordingly. We stress that while small pointing offsets (a fraction of the beam) for each of the 
CO datasets is possible, it is unlikely that instead a single large ($\sim$8\arcsec) pointing offset 
could be present for just one of the datasets (and such a large offset in a single dataset would be highly obvious).
We note that although applying the pointing offsets results in generally better alignment of features in the two 
CO images, it does not significantly affect the overall results presented in this work.

We then convolved all the images to match the PSF of the HARP-B data (15\arcsec\/ FWHM) 
and regridded them onto the same pixel grid as the HARP-B data (7.5\arcsec\/ pixels). 
For the MIPS and IRAC images we obtained convolution kernels for the individual wavebands 
from \citet[][]{gordonetal08}, which are produced using the PSF simulator STinyTim 
\citep[][]{krist02} and empirically determined, respectively. 
Detailed information is given in \citet[][]{gordonetal08}. From these, we created 
kernels that match the resolution of each dataset to the HARP-B \cothree\/ PSF and applied 
these to each dataset.
For the CO and \Hi\/ images we used Gaussian kernels.
\footnote{Kernels are available from http://dirty.as.arizona.edu/$\sim$kgordon/mips/conv\_psfs/conv\_psfs.html}
In our analysis in Sections~\ref{sec:coratio} and~\ref{sec:spiral}, only pixels detected at 
$>$3$\sigma$ in the relevant images are used.

\subsection{The $X$-factor and the conversion of CO line intensity to molecular gas surface density}
\label{sec:method:gas-mass}

The surface density of molecular gas, and related parameters such as the gas-to-dust ratio, 
are dependent on the ubiquitous CO--to--\Htwo\/ conversion factor (`$X$-factor') \xco. 
Application of 
a constant X-factor across a galaxy assumes that there is little variation with metallicity 
or star formation activity. Yet, the metallicity is found to vary with radius in most 
spiral galaxies, leading many authors to suggest that \xco\/ may vary with 
metallicity in spiral galaxies \citep[e.g.][]{arimoto96,israel05}, while other authors suggest that 
\xco\/ may depend on other factors such as the cosmic-ray rate \citep[e.g.][]{bell06}.
For example, \citet{garcia-burilloetal93a} found evidence that the value of \xco\/ 
may be a factor of about two smaller in the inter-arm region compared to the spiral arms.

Moreover, the $X$-factor for conversion from CO line emission to molecular 
gas column density is usually based on the \coone\/ line, and the value of \xco\/ appropriate for 
the \jthree\/ transition of CO is rather uncertain. Previous authors have attempted 
to use \cothree\/ data as an indirect measure of \Htwo, by assuming a \coone\/ 
\xco\/ factor and converting from \cothree\/ intensity to \coone\/ intensity using the 
measured CO~$J$=3-2/$J$=1-0 ratio. However, this method assumes that the 
\xco-factor is constant, and does not vary, for example, 
with galactocentric radius, and that the different CO lines trace the same 
molecular gas. Other authors have derived values for the \cothree--to--\Htwo\/ conversion factor, \xcothree, 
using numerical models, for the central regions of galaxies 
where \ico$>$100\,\kms, and thus it is unclear whether it is applicable across the 
galaxy as a whole, in particular at much large galactocentric radii and in inter-arm 
regions where the CO intensities are considerably lower.

Studies of the \Htwo\/ surface density using the \cotwo\/ and \coone\/ datasets used herein 
have been presented by previous authors (notably, \citet[][]{schuster07} and \citet[][]{nakaietal94}).
In the present work, however, we are concerned with examining the variation of CO properties 
as a function of galactocentric radius and physical location (arm or inter-arm region etc.) 
across \mf, and thus we choose to discuss the distribution
in terms of observed CO integrated intensities and do not attempt to apply a 
\cothree--to--\Htwo\/ conversion factor.  
Also, examining relationships, e.g. the relationship between \cothree\/ and PAH emission 
(see Section~\ref{sec:copah}), in terms of CO intensities rather than molecular gas surface densities 
allows us to make a consistent comparison with previous authors who disussed CO intensities
\citep[e.g.][]{reganetal06,bendo10}.
However, we will use the \cothree\/ intensity to investigate whether the 
CO~\jthree/\jone\/ ratio, or its dependence on another property of the galaxy such as \cothree\/ 
intensity or galactocentric radius, could be used to scale the \coone\/ 
$X$-factor for use with the \cothree\/ line.

\section{A map of \mf\/ in \cothree\/ emission and comparison to other wavebands}
\label{sec:images}

The HARP-B \cothree\/ integrated intensity map of \mf, shown in Figure~\ref{fig:maps-ico32}, 
is the first sensitive CO map of the entire galaxy disk in the \cothree\/ transition.  
The spatial scale is $\sim$600\,pc per 15\arcsec\/ beam, or $\sim$300\,pc per pixel.
The first contour level shown corresponds to 3$\sigma$, where $\sigma\sim$0.17~K\,\kms 
(\tmb). \cothree\/ emission is detected out to a radius of $\sim$12\,kpc, and is detected 
everywhere over a central area of $\sim$14$\times$10\,kpc.

In this section we describe the \cothree\/ image, which shows CO emission associated with 
molecular gas. We also give a qualitative comparison with images of 
other spiral arm tracers: lower CO transition tracers of molecular gas, \Hi\/ emission that 
shows atomic gas, PAH emission (using the 
\pah\/ image produced as described in Section~\ref{sec:ancillary:sings}), \Ha\/ that traces 
HII regions, and MIPS 24\micron\/ emission that traces recent dust enshrouded
star formation (and may also include a component of diffuse 
24\micron\/ emission that is not associated with recent star formation; Section~\ref{sec:spiral}). 
These images are shown at their native resolutions in Figures~\ref{fig:maps-ico32}--\ref{fig:maps-other}, 
with contours of the \cothree\/ emission superimposed.

\subsection{The \cothree\/ image}
\label{sec:images:co32}

\begin{table*}
\centering
\begin{minipage}{140mm}
\caption{Typical values of \cothree\/ line widths and brightness temperatures}
\begin{tabular}{llccc}
\hline
 & Central kpc & Arms & Outer arms & Inter-arm\\
\hline
 Linewidth (\kms) & 40--80 [$>$100] & 20--50 & 15--20 & 15-20 \\
Brightness temperature (K) & 0.3--0.7 [0.8] & 0.1--0.5 & 0.1 & $<$0.1 \\
\hline
\end{tabular}
\label{tab:linewidth+temp}
\end{minipage}
\end{table*}

The \cothree\/ emission traces the two-armed spiral pattern, with one arm unwinding from the 
western side of the nuclear disk (hereafter `Arm I') and the other arm unwinding from 
the eastern side of the nuclear disk and stretching towards the companion galaxy, NGC\,5195, 
located at $\sim$11\,kpc to the northeast of \mf\/ (hereafter `Arm II'). 
The two arms are labelled in Figure~\ref{fig:maps-ico32}. 
Arm I exhibits continuous \cothree\/ 
emission out to a radius $r\sim$250\arcsec\/ ($\sim$10\,kpc), with a notable `kink' to the 
northwest at $r\sim$110\arcsec, where a peak of emission extends into the inter-arm 
region via a spur or 
bridge joining Arm I with Arm II. On the other 
hand, Arm II exhibits depression in the CO emission at $\sim$100\arcsec\/ west. 
This is not a void but 
rather the CO emission here is at a comparable level to that in the inter-arm region. 

Beyond this gap, there is a region of enhanced emission in Arm II at $\sim$120\arcsec\/ 
($\sim$5~kpc) southwest, where the intensity is the highest 
of any emission in the spiral arm outside the central $\sim$2~kpc. This emission peak is 
mirrored symmetrically opposite by such a peak in Arm I at 120\arcsec\/ northeast. 
Enhanced star formation activity \citep[][]{calzetti05} and [CII] emission \citep[][]{nikola01} 
have previously been reported at these two locations, which are near the predicted corotation 
radius \citep[160\arcsec;][]{garcia-burilloetal93a,garcia-burilloetal93b} of the strong density 
wave pattern implied by kinematic studies 
\citep[e.g.][]{aalto99,garcia-burilloetal93a,kuno-nakai97,shetty07}. Enhanced emission 
at these locations is interpreted as being triggered by the interaction with the companion
\citep[e.g.][]{toomre72,nikola01}.
As described previously by a number of authors \citep[e.g.][]{nakaietal94}, 
the locations of these peaks also mark the locations beyond which the emission no longer 
follows a simple (e.g. logarithmic) spiral pattern.

Along the inner part of Arm~I the emission peaks are fairly regularly spaced (at intervals 
of typically $\sim$1.8~kpc), while for Arm~II fewer distinct peaks are discernible.
The emission along the outer parts of both spiral arms appears to be more fragmented, 
as noted for \cotwo\/ emission by \citet{schuster07}. However, the outer part of Arm~I 
to the west exhibits continuous low-level \cothree\/ emission out to $\sim$10~kpc 
(i.e. comparable to the level of inter-arm emission) with emission peaks fairly regularly 
spaced along the arm (at intervals typically of the order of $\sim$2.5\~kpc). 

The arm--inter-arm intensity contrast is high, and the brightest \cothree\/ emission clearly 
follows the spiral structure. 
Nonetheless, inter-arm emission is detected at $>$3$\sigma$ everywhere out to a 
galactocentric radius of $\sim$5~kpc and is detected for some regions out to $\sim$10\,kpc. 
The location of the highest arm/inter-arm contrast is at $\sim$1.7~kpc south-west 
(just to the west of the brightest emission peaks in the image) where 
the intensity drops by a factor of $\sim$14 over just 9\arcsec. 
We will return to the discussion of the spiral structure later in this section and 
in Section~\ref{sec:spiral}.

\begin{figure*}
\centering
\includegraphics[width=17cm, clip]{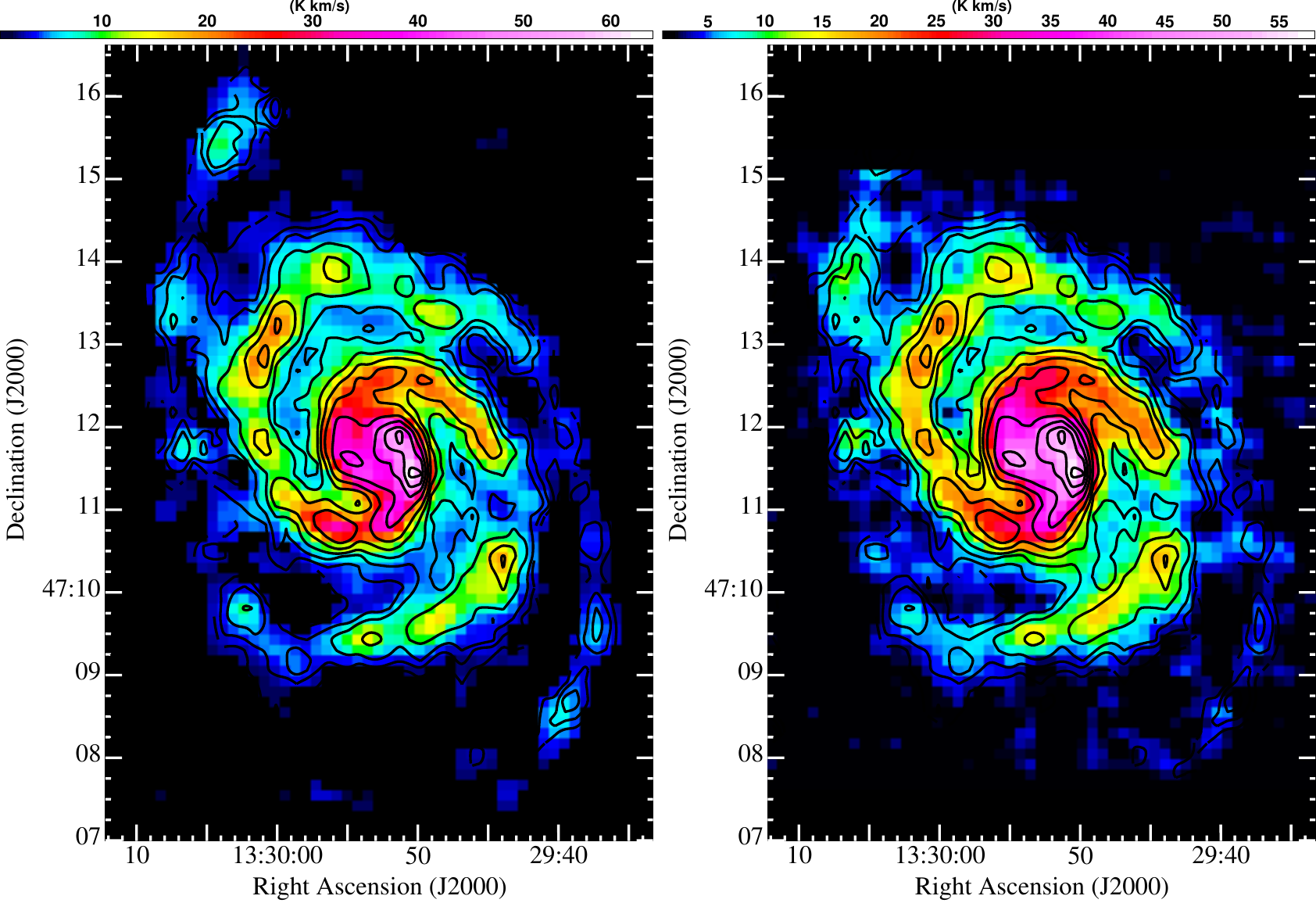}
\caption{\cotwo\/ and \coone\/ integrated intensity images, from \citet{schuster07} 
and  \citet[][]{koda09,koda11}, respectively, shown on logarithmic colour scales.
Pixels where no emission was detected are shown in black.
Images are shown at their native resolution (12\arcsec and 16\arcsec, respectively; 
see ~\ref{sec:obs-data}) but on the same 7.5\arcsec\/ pixel grid as for the HARP image in 
Figure~\ref{fig:maps-ico32}.
Contours show the HARP \cothree\/ integrated intensities as in Fig.~\ref{fig:maps-ico32}.
The FOV of all the images is identical to the HARP image shown in 
Figure~\ref{fig:maps-ico32}; the spatial scale is indicated in Figure~\ref{fig:maps-ico32}.
}
\label{fig:maps-otherco}
\end{figure*}

\begin{figure*}
\centering
\begin{minipage}{170cm}
\includegraphics[width=16cm, clip]{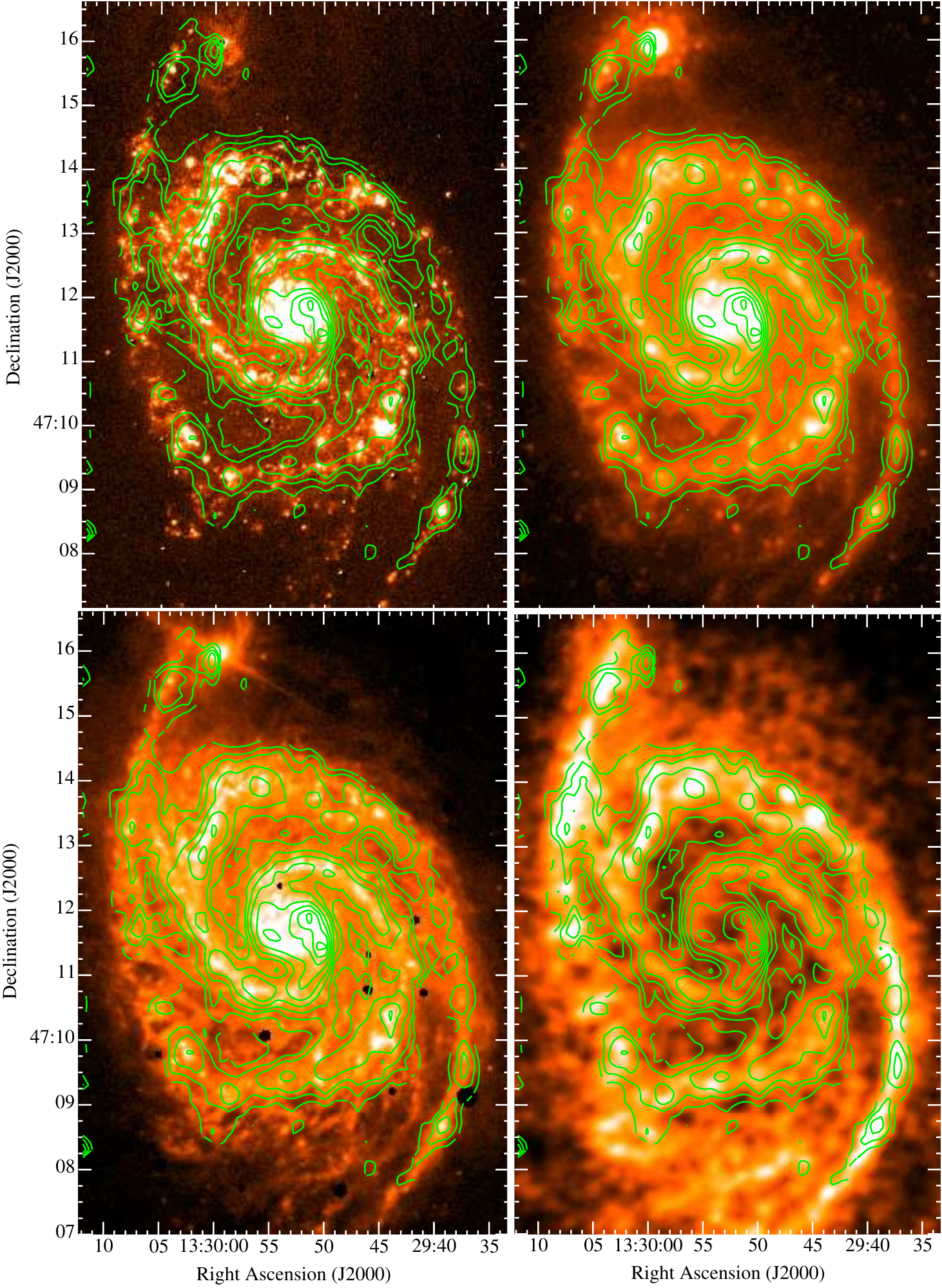}
\end{minipage}
\caption{Images of (top row) \Ha, 24\micron, (bottom row) \pah\/ (produced as described in 
Section~\ref{sec:obs-data}) and \Hi\/ emission.
Colour scales are chosen so as to best demonstrate the galaxy structure in each image 
(logarithmic colour scales for the \pah\/ and 24\micron\/ images).
For the \Hi\/ image the colour scale shows the range from 0--0.3~Jy/beam\,\kms and, 
for the \pah\/ image the colour scale ranges from 0--90~MJy\,sr$^{-1}$ for \mf\/ (the 
companion galaxy NGC\,5195 has higher values).
The images are shown at their native resolutions (Section~\ref{sec:obs-data}; all have higher resolution 
than the CO images in Figures~\ref{fig:maps-ico32} and ~\ref{fig:maps-otherco}). 
Contours show the HARP \cothree\/ integrated intensities as in Figure~\ref{fig:maps-ico32}.
}
\label{fig:maps-other}
\end{figure*}

There are several obvious spurs of \cothree\/ emission connecting the two arms. 
One of the brightest features is to the north-west, at a galactocentric radius of 
$\sim$100\arcsec\/ ($\sim$4~kpc). To the south-west, at radii of $\sim$50\arcsec--110\arcsec\/ 
($\sim$2--4~kpc), there are two further spurs, while another strong feature is seen to 
the north-east at a radius of $\sim$150\arcsec\/ ($\sim$6\,kpc). 
These spur-like features are also very apparent in the \cotwo, \coone\/ and 
\Hi\/ images, the latter feature being especially apparent in the \Hi, \Ha, 24\micron, 
and \pah\/ images (see Figures~\ref{fig:maps-otherco} and~\ref{fig:maps-other}). 
To the south, there is a hint of two faint spurs of emission connecting the two arms at 
$\sim$3\arcmin--4\arcmin\/ 
(7--10\,kpc). Although in these regions the CO emission is detected in typically only a handful of 
channels, with the signal-to-noise over these channels ranging from 2--7$\sigma$,
we note that peaks in the stronger of the two spurs are also apparent in the \cotwo\/ image 
(see Figure~\ref{fig:maps-otherco}) and
both features are clearly detected in recent {\it Herschel} SPIRE 250\micron\/ and 350\micron\/ 
observations \citep{mentuchcooper12} which have comparable resolution. The \cothree\/ image is seen to 
closely resemble the 250\micron\/ SPIRE image.

Typical values of \cothree\/ line widths and brightness temperatures at different locations in the galaxy are 
given in Table~\ref{tab:linewidth+temp}.
As expected due to beam smearing, the broadest  line spectra are found within the central 
kiloparsec (central $\sim$25\arcsec) 
of the galaxy, with typical widths of $\sim$40--80~\kms, and $>$100~\kms\/ at the very centre.
Here, brightness temperatures are found to be typically 0.3--0.7~K, with the highest values ($\sim$0.8~K) 
found not right at the centre but corresponding to the locations where the spiral arms begin to wind out. 
The brightness temperature in the spiral arms is typically found to range from 0.1--0.5~K, with 
no obvious dependence on radius until the outermost region of the arms, where values are 
$\sim$0.1~K. Typical line widths in the spiral arms, 
however, are found to decrease with distance along the spiral 
arms, from $\sim$50~\kms\/ to $\sim$20~\kms, as has previously been noted for \cotwo\/ by 
\citet{garcia-burilloetal93a} and \citet[][]{hitschfeld09} who attribute the decrease 
in line width to streaming motions \citep[e.g.][]{garcia-burilloetal93b}. 
In the outer spiral arms line widths are typically 15-20~\kms, 
comparable to the widths found in the outer disk and in the inter-arm regions. 
Brightness temperatures in the inter-arm regions are found to be $<$0.1~K. Overall, \cothree\/  
line widths in the centre of the galaxy and in the spiral arms are in very good agreement with 
those found for \coone\/ by \citet[][]{nakaietal94} and for \cotwo\/ by 
\citet{garcia-burilloetal93a} and \citet[][]{hitschfeld09}. 

\subsection{Comparison to \cotwo\/ and \coone\/ images}
\label{sec:images:otherco}

CO~J=2-1 and \coone\/ integrated intensity images, at their native resolutions, are shown  
in Figure~\ref{fig:maps-otherco}.
All images are convolved to a common 
scale of $\sim$600\,pc and gridded on to the same 7.5\arcsec\/ pixel grid as the HARP data, 
as described in Section~\ref{sec:obs-data}.  
The \cotwo\/ and 
\coone\/ emission is seen to trace the same two-armed spiral pattern as described for 
\cothree. 

The \cotwo\/ emission in particular closely traces the same features described 
above for \cothree, with clear distinction between the spiral arm and inter-arm regions. 
In fact, 
the peaks of \cothree\/ and \cotwo\/ emission are remarkably well aligned; at this spatial 
resolution we find no significant emission peaks in either map that are not in excellent 
agreement. Even in the central region where, at $\sim$50\arcsec\/, the spiral arms begin 
to wind out and Arm I is characterised by two peaks of \cothree\/ emission, these two 
emission peaks are seen at the same spatial location in the \cotwo\/ image.
All the bright peaks of \cothree\/ emission appear to correspond to peaks 
in the \cotwo\/ image and most of the bright \cothree\/ peaks appear to 
have corresponding peaks in the \coone\/ image. 
Conversely, we find no bright peaks in the \cotwo\/ and \coone\/ images that 
are not the also the locations of bright peaks in the \cothree\/ image.

The highest arm/inter-arm contrast in the \cotwo\/ image is found at the same location as 
for the \cothree\/ image, i.e. $\sim$1.7~kpc south-west of the centre, 
although the \cotwo\/ integrated intensity drops by only a factor of about six over 
9\arcsec, and falls by a factor of 
14 over $\sim$16\arcsec. \citet{garcia-burilloetal93a} found a similar result at this location 
using a different \cotwo\/ dataset. These authors point out that the sudden drop between the 
spiral arm and inter-arm may in part be related to the fall-off of the nuclear component.

The \coone\/ emission appears more diffuse with fewer apparent features. For example, the 
central \coone\/ emission is characterised by a single peak that is located between 
the peaks seen for the higher transitions. Indeed, a fundamental difference between \coone\/ 
and the higher transitions is to be expected, since the \coone\/ line 
traces clouds of molecular gas, for example relatively low density cold diffuse gas, that are not 
(or only marginally) traced by the higher CO transitions.

For all three CO transitions the spiral arms show fairly regularly spaced concentrations or ``clumps'' of 
molecular gas, and, as noted for the lower CO transitions by previous authors 
\citep[e.g.][]{garcia-burilloetal93a,schuster07}, at a spatial resolution of $\sim$600~pc may 
be referred to as GMAs and may represent either bound clusters or random superpositions 
of GMCs \citep[e.g.][]{rand-kulkarni90,garcia-burilloetal93a}. 
The outer parts of the two arms 
appear more fragmented than the inner parts for all three CO transitions.

\begin{figure*}
\centering
\rotatebox{0}{\includegraphics[width=15cm, clip]{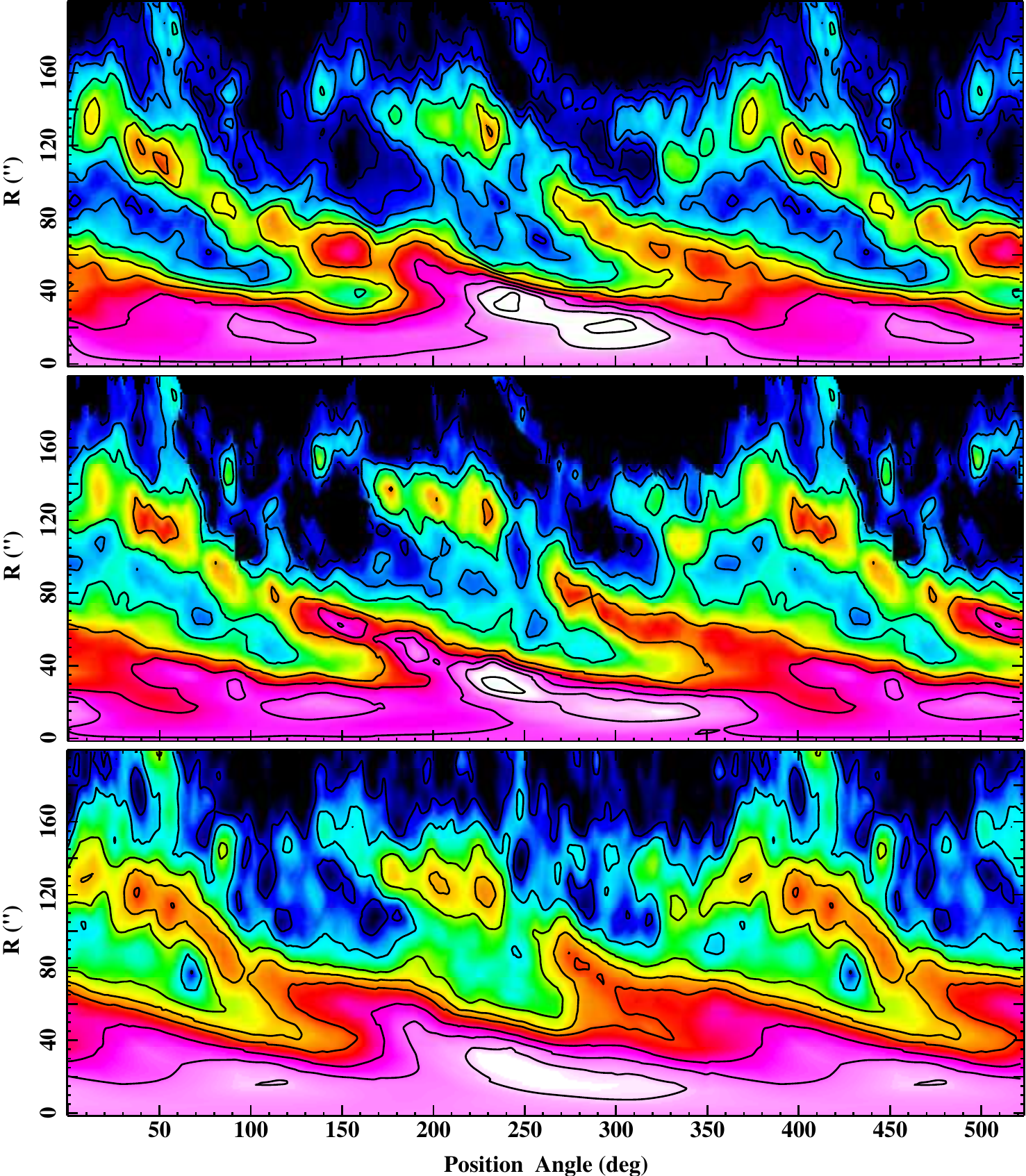}}\\
\caption{Maps of \cothree\/ (HARP-B, this work), \cotwo\/ and \coone\/  integrated intensities 
shown in polar coordinates, where axes show galactocentric radius (in arcseconds; 1\arcsec$\sim$40~pc) 
and position angle. The images have been de-projected using the parameters in Table~\ref{properties} 
as described in Section~\ref{sec:images:otherwave}.
The images are shown on the same colour scales as in Figures~\ref{fig:maps-ico32} and~\ref{fig:maps-otherco}.
Contours for the \cothree\/ image are the same as in Figure~\ref{fig:maps-ico32}. 
For the \coone\/ image contours are at 2, 5, 10, 15, 20, 30, 40 and 50~K\,\kms\/ and for the 
\cotwo\/ image contours are at 1, 3, 7, 18, 30, 41, 52, 64 ~K\,\kms (\tmb). 
Arm I begins at P.A.$\sim$190$\degr$ and Arm II begins at P.A.$\sim$370$\degr$.
}
\label{fig:maps-polar-co}
\end{figure*}

\begin{figure*}
\centering
\rotatebox{0}{\includegraphics[width=15cm, clip]{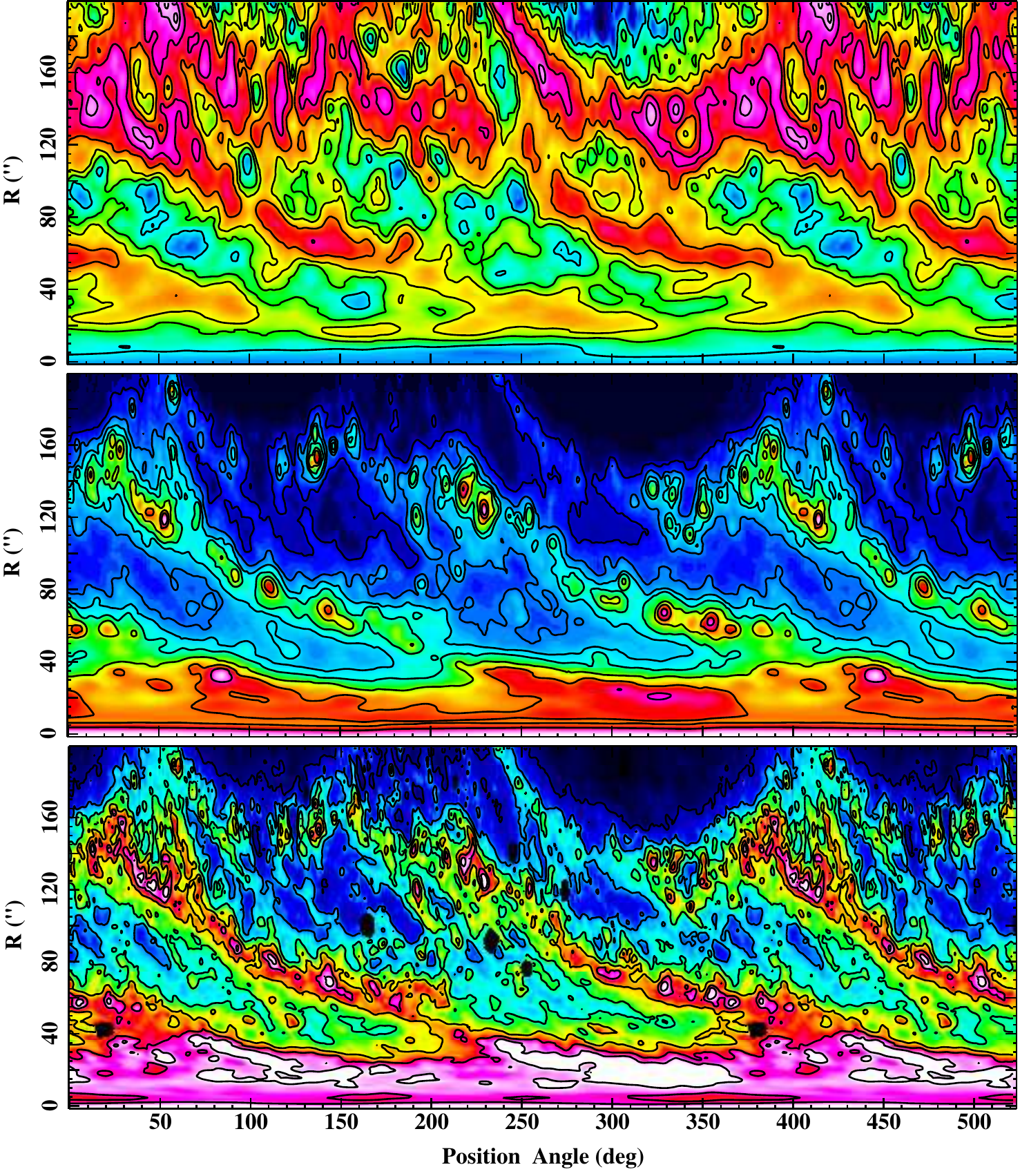}}\\
\caption{As for Figure~\ref{fig:maps-polar-co} but for \Hi, 24\micron\/ and \pah\/ emission (from top).
\Hi\/ contours are at 36, 54,72, 108, 162, 234, and 288~mJy/beam~\kms.
24\micron\/ contours are at 1.5, 3, 5, 8, 15, 30, 65~MJy/sr.
\pah\/ contours are at 0.5, 1.5, 3, 6, 12, 23~MJy/sr.
}
\label{fig:maps-polar-other}
\end{figure*}

\subsection{Comparison to other wavebands}
\label{sec:images:otherwave}

The \Ha, 24\micron, \pah\/ and \Hi\/ images are shown, at their native resolutions, 
in Figure~\ref{fig:maps-other}.
The \pah\/ image clearly traces the clumpy dust structures of the ISM. 
The \Hi\/ emission is stronger in the outer regions, and only a faint spiral structure is seen 
in the inner regions, as is typical in the inner parts of late type spirals 
where star formation is more active and the molecular phase dominates. By simple inspection of 
the images it is clear that the intensities of CO and \Hi\/ are anti-correlated 
(also see Section~\ref{sec:radial-variation}).

Compared to the \cothree\/ emission, \Ha, \Hi, \pah, and 24\,\micron\/ peaks in the 
northern parts of both 
spiral arms (within the co-rotation radius) are found to be slightly offset, 
by 300--400\,pc (7\arcsec--10\arcsec), towards the convex (downstream)
side of the arms. This is particularly clear for Arm~II at $\sim$2.5~kpc north.
This is in agreement with the findings of previous authors 
\citep[e.g.][]{tilanus-allen91,garcia-burilloetal93a,kunoetal95,bastianetal05,egusa09,louie13}, 
although we note that \cothree\/ (and likewise the \cotwo\/ and \coone) emission 
is nonetheless present at the offset locations. 
Elsewhere in the spiral arms, even in the outer arms up to $\sim$10~kpc,  
we see excellent agreement between bright peaks of \cothree\/ 
emission and bright peaks in the \Hi, \pah, \Ha\/ and 24\micron\/ images.
One clear difference is that in the outer spiral arms of the \cothree\/ (and likewise \cotwo) image, 
despite the good agreement between bright peaks we note a tendency for the 
\Hi, \pah\/ and 24\micron\/ emission to be more extended on the convex side of the arms. 
This is particularly apparent at the western and north-eastern sides of Arm I, close to 
locations of spurs of emission bridging the region between the inner spiral structure and the 
outer arms that are particularly clear in the \pah\/ image.
 
Comparison to the \cotwo\/ image from \citet{schuster07} presents a very similar result, 
since the \cothree\/ and \cotwo\/ peaks are 
coincident, and we find excellent agreement between bright peaks of \cotwo\/ emission and all the 
other emissions even in the outer part of Arm I where these authors noted 
a slight offset. The relative locations of peaks in the \cotwo\/ and \Ha\/ images are in 
excellent agreement with those presented for different \cotwo\/ and \Ha\/ datasets by 
\citet{garcia-burilloetal93a}, and thus we believe that the slight offset noted by 
\citet{schuster07} was likely due to a small pointing offset for which we have 
corrected in the present work (see Section~\ref{sec:obs-data} for details of our treatment 
of the datasets).

The \cothree\/ image also shows very good correspondence with with the \pah\/ image, with 
both images appearing to trace the same spiral spiral pattern. The locations of most peaks 
(and likewise troughs) in the \cothree\/ image are also the locations of peaks and troughs 
in the \pah\/ image. This is particularly obvious in the outer spiral arms and at the location 
of the spurs of emission spanning the inter-arm region to the southwest. 
The converse is also true, i.e. most features in the \pah\/ image have counterparts in the 
\cothree\/ image.
There are, however, 
some clear differences. For example, to the south, where Arm I begins to unwind, there is a 
dip in the \pah\/ surface brightness that is not seen in the \cothree\/ image. Also, in Arm II 
at $\sim$5.5~kpc the \cothree\/ peak is coincident with a peak in the \pah\/ image, but there 
is a second \pah\/ peak with no corresponding \cothree\/ feature. The distribution of emission 
in the \cothree\/ and 24\micron\/ images is also generally similar, but the differences described 
for the \pah\/ image are also seen for the 24\micron\/ image. One notable difference between the 
\cothree\/ and 24\micron\/ images can be seen for the outer western part of Arm I, 
where the emission appears very smooth at 24\micron\/ while several \cothree\/ knots are 
seen along the arm.
We will return the the 
subject of the relationship between the \cothree\/, \pah\/ and 24\,\micron\/ 
images in Section~\ref{sec:spiral}.

Many of the same features can be seen across most wave bands, despite the different native 
resolutions. To allow a clearer visualisaion of the comparison, and to aid the eye in 
delineating spiral features, in Figures~\ref{fig:maps-polar-co} and~\ref{fig:maps-polar-other} 
we display the CO, \Hi, \pah\/ and 24\micron\/ images in polar coordinates. 
These images have first been de-projected using the inclination angle, position angle and 
central position given in Table~\ref{properties}.

In polar coordinates, Arm I can be seen to emerge from a radius of $\sim$50\arcsec\/ 
in the central region at a position angle of $\sim$190$\degr$. Arm II begins at $\sim$370$\degr$ (i.e. 10$\degr$).
Many of the features of the CO emission described in Section~\ref{sec:images} are 
clearly emphasized in the polar images, and it is particularly striking how well-defined the 
spiral arm structure is in \cothree\/ and \cotwo\/ emission. 
Conversely, the \coone\/ image appears considerably more diffuse, with clear emission 
in the inter-arm regions. There are similarities, however. Spurs of emission 
connecting the two arms can clearly be seen in all three CO images at position angles of 
$\sim$60$\degr$ and $\sim$340$\degr$, 
and the point at which the spiral pattern deviates from a simple
logarithmic spiral is clearly discernible at position angles of $\sim$50$\degr$ and $\sim$230$\degr$. 
The latter are  the locations of the symmetrically opposite bright peaks in 
each of the spiral arms described in Section~\ref{sec:images:co32}.

\subsection{Contamination of submillimetre continuum by \cothree\/ line emission}
\label{sec:images:contam}

\begin{figure}
\centering
\includegraphics[width=8.5cm, clip]{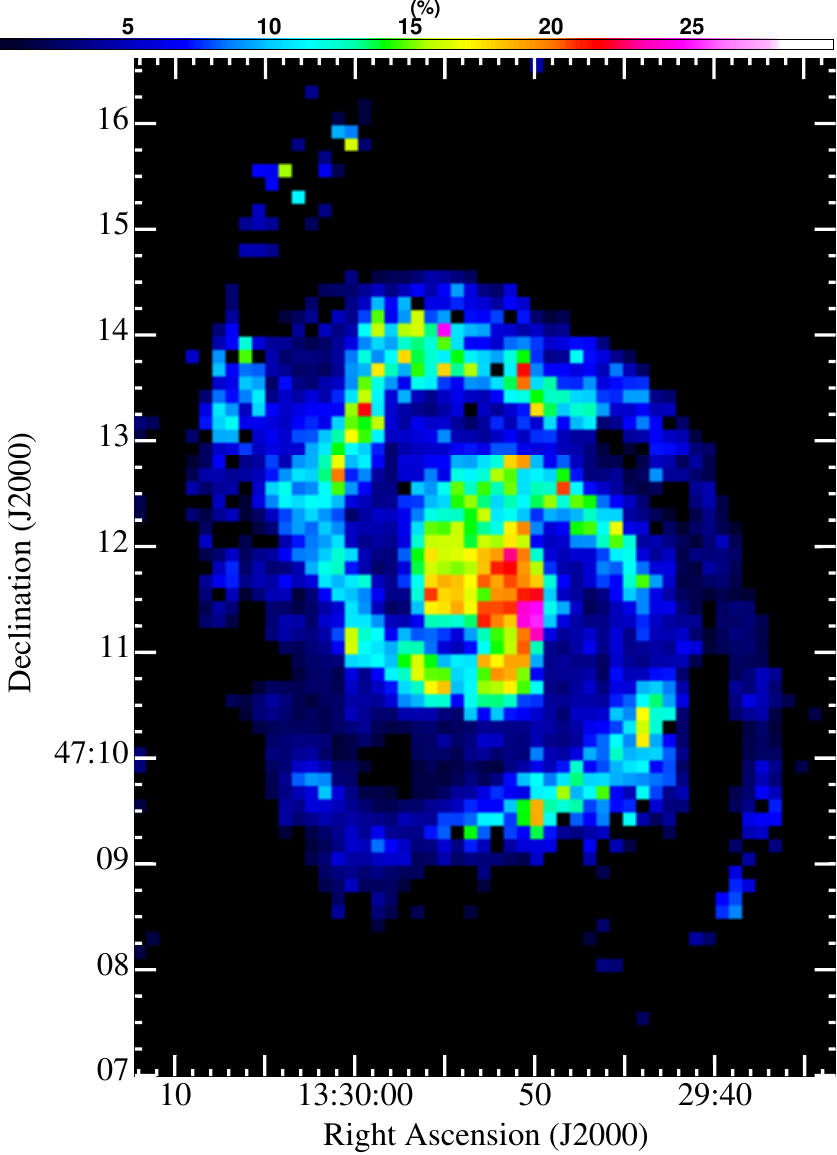}
\caption{Percentage contamination at 850\micron\/ by \cothree\/ line emission, calculated as described 
in Section~\ref{sec:images:contam}. Only pixels detected at $>$1.5$\sigma$ in the \cothree\/ and 850\micron\/ images are 
considered. Values are typically 10--20\% in the spiral arms and are highest in the central 
$\sim$2\,kpc radius (on average $\sim$20\% but reaching a maximum of 26\% in the inner part of Arm II). 
In the inter-arm regions and outer spiral arms values are $<$10\%. 
}
\label{fig:contam}
\end{figure}

Submm continuum observations of Galactic and extragalactic objects may suffer from contamination 
from inband line emission, which for many sources is primarily \cothree\/ line emission, 
due to the broad passband of many submm instruments such as SCUBA; the centre of the SCUBA 
850\micron\/ bandpass is very close to the rest frequency of the \cothree\/ transition (345.796\,GHz). 
Continuum 850\micron\/ emission is typically used to trace cool dust, and since correcting 
850\micron\/ 
emission for the contribution from CO will lead to a higher dust temperature and lower dust mass, 
knowledge of the amount of contamination can be vital. In particular, detailed knowledge of the 
amount of contamination by the \cothree\/ line in a nearby galaxy such as \mf, where we can 
investigate the percentage contamination as a function of radius and spiral structure, 
could be extremely useful for future galaxy studies.

We carried out this investigation using our \cothree\/ map together with the total 
SCUBA 850\micron\/ map from \citet[][]{meijerink05}, which includes emission from an 
exponential dust disk 
\citep[see][for details and for a description of the data reduction procedure]{meijerink05}. 
Since both the \cothree\/ and 850\micron\/ maps were obtained at the JCMT the beam sizes are 
comparable so we can compare the two maps directly and the only treatment of the 850\micron\/ 
map was to regrid it onto the same pixel grid as the \cothree\/ map. 
Only pixels detected at $>$1.5$\sigma$ in both the \cothree\/ and 850\micron\/ images were considered.

We first produced an image of the SCUBA equivalent flux produced by the \cothree\/ line at the recessional 
velocity of \mf\/ using the profile of the SCUBA filter bandpass and a 15\arcsec\/ beam size.
We then divided this image by the SCUBA 850\micron\/ image to produce an image of the percentage 
contamination at 850\micron\/ by inband \cothree\/ emission. This is shown in Figure~\ref{fig:contam}. 
Values are typically 10--20\% in the spiral arms and are highest in the central $\sim$2\,kpc radius 
(on average $\sim$20\% but reaching a maximum of 26\% in the inner part of Arm II). In the inter-arm 
regions and outer spiral arms values are $<$10\%. The average over the disk is ~9\%. We note, however, that 
if we perform the same analysis using the exponential disk-subtracted 850\micron\/ map from 
\citet[][]{meijerink05} the percentage contamination in the spiral arms is considerably higher.

\section{The geometry of central molecular emission}
\label{sec:geometry}

\begin{figure*}
\centering
\includegraphics[width=16cm, clip=true, trim=0.5cm 0 0 0]{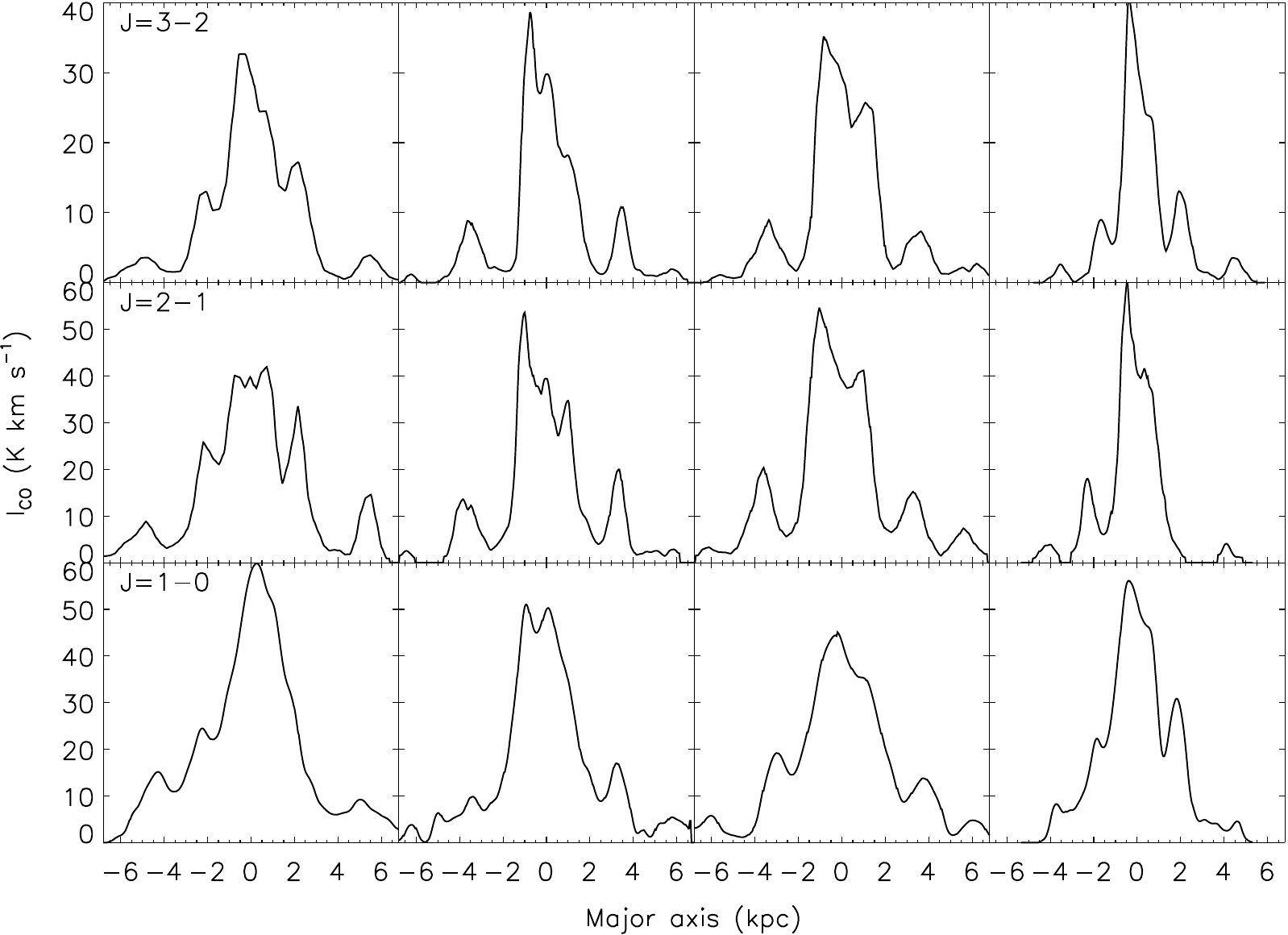}
\caption{Profiles of CO integrated intensity (\tmb) along the major axis of \mf\/  when projected to 
various edge-on orientations (from left: major axis along ${\rm P.A.}=0, 45, 90~{\rm and}\/ 135$ degrees), using the parameters in Table~\ref{properties}.  Rows show the different CO transitions as indicated. }
\label{fig:edge-on}
\end{figure*}

Several authors have attempted to describe the geometry of the distribution of molecular 
gas in the central regions of galaxies that are more highly inclined than \mf\/ 
based on the observed profile 
of the central gas component (often a `two-horned' profile), leading to the conclusion that the gas is located 
in a molecular ring \citep[e.g.][]{irwin10}. However, the inclination of a galaxy may 
determine how reliably the geometry of its gas distribution can be determined. If, for 
example, a galaxy has bright molecular gas rich spiral arms, as is seen for \mf, this 
could potentially mimic a ring if observed edge-on \citep[c.f.][]{israeletal06}. 
We can therefore use the \mf\/ 
\cothree, \cotwo\/ and \coone\/ maps as a simple test of this scenario, by projecting 
the CO integrated intensity images to several different edge-on orientations and examining 
the shape of the profiles along the major axis. 

Figure~\ref{fig:edge-on} shows major axis profiles of \mf\/ for a major axis oriented along 
$P.A. = 0, 45,$ 90 and 135$\degr$, produced using the 
parameters given in Table~\ref{properties}.
It can clearly be seen that the shape of the \cothree\/ and \cotwo\/ profiles at each 
projected orientation are similar. However, the important thing to note is that 
it is also clear that the shape of 
the edge-on profile can appear `centrally-peaked' or `two-horned' (see the 
second column of Figure~\ref{fig:edge-on} in particular) depending on the 
orientation. Thus it appears that, depending on viewing angle, a galaxy like \mf\/ with bright 
molecular gas rich inner spiral arms may be observed to exhibit a two-horned profile that mimics 
a molecular ring.

\section{CO line ratios}
\label{sec:coratio}

\begin{figure*}
\centering
\includegraphics[width=\textwidth, clip]{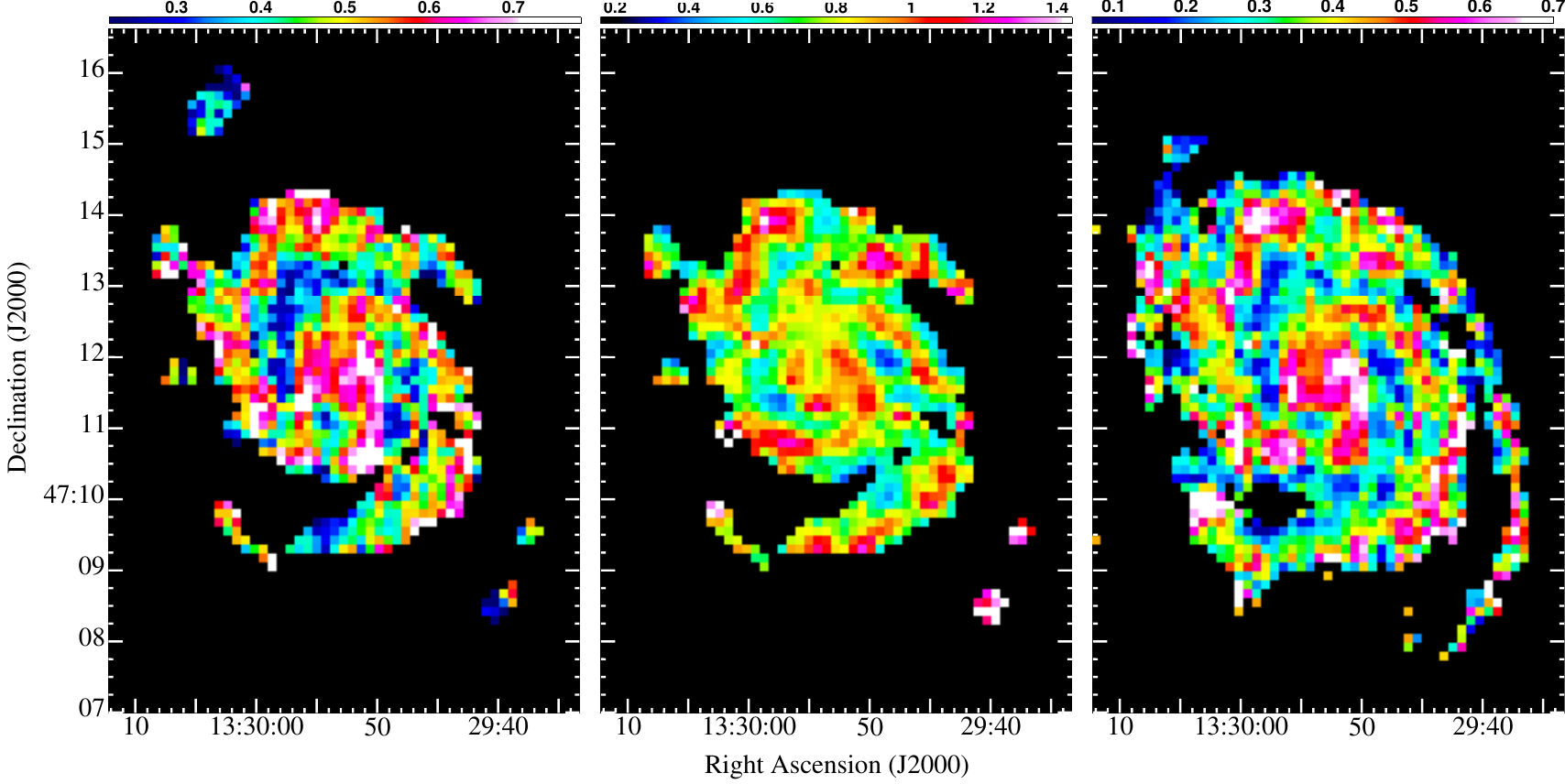}
\caption{Maps of the ratio of the  J=3-2/J=2-1,  J=2-1/J=1-0, and J=3-2/J=1-0 CO integrated 
intensities (from left), shown on colour scales that are centred on the median ratio value in each map 
and have a range of 2$\sigma$ from the median.
Median ratio values are 0.5, 0.80 and 0.36, respectively (see Table~\ref{tab:coratios}).
The maps are produced as described in Section~\ref{sec:coratio}; 
the resolution, pixel size and FOV of each input image has been matched to those of the HARP \cothree\/ 
image (Figure~\ref{fig:maps-ico32}), and only pixels with emission $>$3$\sigma$ in both input images are shown. 
The uncertainties in the CO ratios are mainly dependent on the calibration uncertainties for the 
individual CO images.}
\label{fig:maps-coratio}
\end{figure*}

Maps of the ratio of CO~$J$=3-2/J=2-1 CO~$J$=2-1/1-0 and CO~$J$=3-2/J=1-0 integrated 
intensities are shown in Figure~\ref{fig:maps-coratio} (hereafter \Rthtw. \Rtwon\/ and \Rthon). 
These CO ratio maps were produced from the CO images shown in Figures~\ref{fig:maps-ico32} 
and~\ref{fig:maps-otherco} 
for pixels where the emission is detected above 3$\sigma$ in both the input images.
Input images (which are on the main beam temperature scale) 
were first matched to the HARP \cothree\/ resolution and pixel grid as 
described in Section~\ref{sec:obs-data}.

The CO ratio maps clearly exhibit a spiral-like structure, with the highest ratio values 
typically located in regions corresponding to the spiral arms and the lowest values 
typically located in the inter-arm regions. Values of the CO ratio in different regions of 
the galaxy are summarised in Table~\ref{tab:coratios}.

We find a median \Rthtw\/ ratio of 0.5 over the whole disk. In the central kiloparsec values 
are somewhat higher, ranging from 0.6 to 0.7 at the central position and towards the 
location where Arm I begins to wind out.
The ratio \Rthtw\/ clearly varies across the galaxy in a way that is similar to the spiral 
structure, with the highest values ($\sim$unity) found in the spiral arms and the lowest 
values ($\sim$0.1) found in the inter-arm regions. 
For \Rthon\/ the median value is 0.36 over the galaxy disk and 
somewhat higher in the central kiloparsec.

For the ratio of CO J=2-1/J=1-0 (\Rtwon) we find a median value of 0.80 over the galaxy 
and somewhat higher in the central kiloparsec. Similar results using the same datasets were presented 
by \citet{koda12}. As for \Rthtw, the ratio \Rtwon\/ also appears to vary in a 
way that follows the spiral structure. However, the main difference is that, for \Rtwon, 
the contrast between the spiral arms and inter-arm regions appears less pronounced over the 
central few kiloparsecs and the highest values are found in the spiral arms at larger 
galactocentric radii ($>$4\,kpc). While values of \Rtwon\/ at larger radii are found to be 
up to a factor of two higher than in the more inner spiral arms, values of \Rthtw\/ in the 
spiral arms appear to be similar at all radii. 
We note, however, that very high or low ratio values found in gradient locations such as at the edges 
of the spiral arms may be unreliable since, despite the care taken to align the maps and convolve them 
to a common beamsize (Section~\ref{sec:images:alignment}), there may still be minute alignment errors 
between the maps or minute differences in matched beamsizes.

\begin{table*}
\centering
\begin{minipage}{140mm}
\caption{CO line ratios in \mf}
\begin{tabular}{lcccc}
 & Global &  Central kpc & Arm & Inter-arm \\
& (1) & (2) & (3) & (4)\\
\hline
 CO J=3-2/J=2-1 & 0.50$^{\dagger}$ $\pm$ 0.14$^{\ast}$ [0.1--1.0] & 0.63 & 0.5 & 0.4 \\
CO J=2-1/J=1-0 & 0.80$^{\S}$ $\pm$ 0.32$^{\ast}$ [0.3--3.1] & 0.86 & 0.9 & 0.5\\
CO J=3-2/J=1-0 & 0.36$^{\ddag}$ $\pm$ 0.17$^{\ast}$ [$<$0.1--1.7] & 0.54 & 0.4 & 0.2\\
\hline
\end{tabular}\\
(1)-(2) Median CO ratio over the full galaxy disk and central kiloparsec, for pixels where the emission is detected above 3$\sigma$ in both input images; the uncertainty indicates the standard deviation of the ratio over the whole disk.\\
(3)-(4) Typical CO ratios in the spiral arms and inter-arm regions.\\
$\dagger$The median \cothree\/ line intensity is 3.6\,K\,\kms and 
the median \cotwo\/ line intensity is 7.5\,K\,\kms.\\
$\ast$The standard deviation of the CO line intensities in the input images is 6, 10 and 5 
for \cothree, \cotwo\/ and \coone, respectively. 
The sources of uncertainty on the CO line intensities are described in Section~\ref{sec:obs-data}.\\
$\S$The median \cotwo\/ and \coone\/ line intensities are 7.6 and 5.5\,K\,\kms, respectively.\\
$\ddag$The median \cothree\/ and \coone\/ line intensities are  2.2 and 3.7\,K\,\kms, respectively.
\label{tab:coratios}
\end{minipage}
\end{table*}

From inspection of Figure~\ref{fig:maps-coratio}, while all three ratio maps appear to 
trace the spiral-arm structure, ratios with \coone\/ appear to show a clearer spiral pattern 
than \Rthtw, which appears somewhat smoother. This is not unexpected -- since the higher CO transitions 
trace the spiral structure better than \coone, ratios of these transitions with \coone\/ will show 
a clear spiral pattern, while if the higher CO transitions trace the spiral structure roughly as well 
as each other then the ratio of the two will produce a map that is somewhat smoother in appearance. 
We discuss in more detail the variation of the CO ratios as a function of distance along the spiral 
arms and inter-arm regions in Section~\ref{sec:spiral}.

We find the values of \Rthtw\/ to be in agreement with the range of values found for 
\mf\/ by previous authors and for spiral galaxies in general 
\citep[e.g.][]{mauersberger99}. 
The average \Rthon\/ line ratio over the whole disk of \mf ($\sim$0.36) falls within 
the range of ratios found by \citet[][0.2--0.7]{mauersberger99}, and is consistent 
with the average values found in a number of other recent \cothree\/ studies of nearby 
galaxies with the HARP-B instrument \citep[e.g.][]{wilson09, warren10}. These authors point out that if 
galaxies have enhanced line ratios in their centres, then previous observations such as those 
by \citet[][]{mauersberger99} that only observed the central regions of galaxies may have 
overestimated the ratios averaged over the whole galaxy disk. 
The factor of $>$10 variation in \Rthon\/ across the disk of \mf\/ is high compared 
to some previous observations \citep[e.g.][]{thornleywilson94,mauersberger99} but is in 
good agreement with the variation found in recent studies by \citet[][]{bendo10} and 
\citet[][]{wilson09} for 
other nearby galaxies observed in \cothree\/ with HARP-B. As these authors 
suggest, a large variation in \Rthon\/ may be real but was not observed before due to some 
regions being undetected or unresolved.

In the present work, no attempt has been made to extract molecular gas parameters from the
measured line ratios, because the observed $^{12}$CO transitions are
strongly degenerate with respect to molecular gas temperature and
density.  This degeneracy can be broken by e.g. observations of an
optically thin isotope such as $^{13}$CO, but these are not available
at present on a galaxy-wide scale. Relevant discussions of a few
individual pointings may be found in \citet{kramer05} and \citet{israeletal06}.

The nuclear region of \mf\/ appears to have line ratios that are different from 
those of much of the disk, indicating perhaps an excitation mechanism
substantially different from the presumed PDR excitation of the disk
gas \citep[also see][]{israeletal06}.  Whether or not this related to the
low-luminosity AGN referred to in the introduction is not presently
clear.  To answer this question reqwuires a more detailed study of the
molecular line emission from the central region.

\section{Radial properties of \mf\/ as a function of the CO spiral structure}
\label{sec:spiral}

In this section we investigate the properties of \mf\/ as a function of galactocentric 
radius and, specifically, as a function of the distance along the observed molecular 
spiral structure.

\subsection{Describing the CO spiral structure}
\label{sec:spiral-method}

The commonly used method of obtaining radial profiles by averaging in elliptical 
annuli over the whole disk is likely to average out the true 
variation of properties along spiral arms and in inter-arm regions, so here we describe our 
alternative approach of producing separate radial profiles of the spiral arms and inter-arm regions.

The spiral pattern at small galactocentric radii 
(R$<$120\arcsec) in \mf\/ has often been described with simple symmetrical logarithmic spirals 
\citep[see e.g.][]{nakaietal94,shetty07}. However, the spiral arm pattern clearly no longer follows 
a logarithmic spiral at R$>$120\arcsec\/ ($>$5~kpc), 
with the fracture in the spiral pattern at R$\sim$120\arcsec\/ and the distortion at radii beyond 
this presumably due to the tidal interaction with the companion galaxy; 
in the images in Figures~\ref{fig:maps-polar-co} and~\ref{fig:maps-polar-other} 
this can be easily seen at P.A.$\sim$50$\degr$ for Arm~I 
and P.A.$\sim$230$\degr$ for Arm~II, 
for all three CO transitions as well as \pah\/ and 24\micron\/ emission. 
In previous studies tracing the spiral structure in \mf\/ to radii much larger than 
120\arcsec, authors have adopted a variety of approaches for defining the spiral 
structure, but have often involved selecting regions based on CO surface brightness. 
For example, \citet[][]{hitschfeld09} attributed all emission above a certain 
threshold molecular gas surface density to 
spiral arms and the remainder to the inner-arm regions, and \citet{foyleetal10} followed a similar 
approach but instead used the surface density of stellar 3.6\micron\/ emission.
This approach, however, has the effect 
of including emission features (spurs) bridging the gap between the spiral arms as part of the 
arms themselves. 

\begin{figure}
\centering
\includegraphics[width=8.2cm, clip]{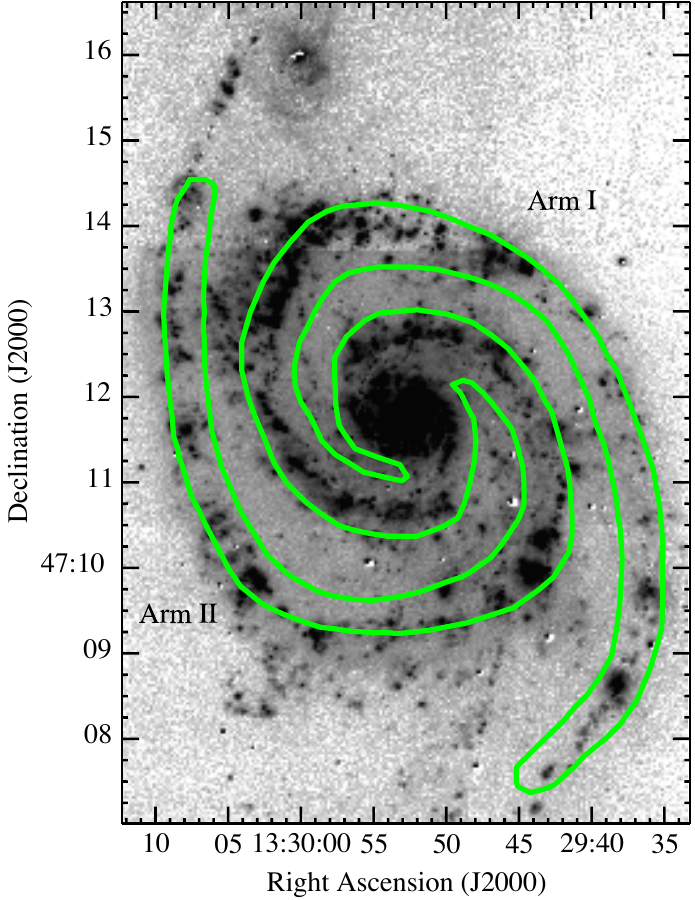}
\caption{Mask used to describe the spiral structure (i.e. used to differentiate arm and inter-arm regions) in \mf\/ (shown in green), derived from the CO (J=3-2 and J=2-1) datasets as described in Section~\ref{sec:spiral}. The greyscale image shows an \Ha\/ image projected to a face-on orientation using the parameters in Table~\ref{properties}.}
\label{fig:spiral}
\end{figure}

In the present work, we instead wish to examine the properties \mf\/ as a function of physical 
location and distance along the {\em observed} spiral morphology out to where 
the interaction with the companion NGC~5195 is taking place ($\sim$11~kpc), 
and thus we require a different approach.
We take the simple approach of defining a mask by eye that broadly describes the 
observed appearance of the CO spiral structure out to large radii -- 
i.e. two spiral arms and two inter-arm regions, where we define the ``spiral arm'' 
region as a continuous broad structure that is 
linked seamlessly from the nuclear region in a broadly spiral pattern. 
The remaining region in between the spiral arms we refer to as the ``inter-arm'' region, 
irrespective of any emission features it may contain. 
The resulting mask -- defined based on the \cothree\/ image while ensuring that it 
adequately describes the spiral structure observed in the other two CO transitions -- 
is shown in Figure~\ref{fig:spiral}.
The width of the spiral arms defined in this way is 
$\sim$1~kpc ($\sim$20\arcsec--30\arcsec), when corrected for the telescope beam, 
which is in excellent agreement with the values found 
for \cotwo\/ and \coone\/ by \citet{garcia-burilloetal93a} and \citet{nakaietal94}, respectively.

We produce radial profiles by plotting the pixels in each spiral arm or inter-arm 
as a function of galactocentric distance (where the central position is defined as given in 
Table~\ref{properties}). These are shown by the black points in 
Figures~\ref{fig:spiral-profile} (filled and open points show Arms~I and II, 
respectively) and \ref{fig:coratio-profile}.  
We also bin this data into 0.5~kpc bins (approximately the size of the HARP-B 
beam-size), and this is shown as the red (Arm I) and blue (Arm II) points in the figures. 
The latter is 
effectively the equivalent of averaging the emission in a series of circular annuli 
separated by 0.5~kpc in which {\it only detected pixels} are considered. 
We do not set undetected pixels to zero since there is considerable difference in the 
sensitivity of the three CO maps we wish to compare (see Section~\ref{sec:obs-data}). 
We note that the CO profiles can be reliably interpreted out to $\sim$5--6~kpc, beyond 
which the number of detected pixels 
is relatively small.

All radial profiles were produced using de-projected images produced as in 
Section~\ref{sec:images:otherwave} and that emission within the central 1.7~kpc is not 
considered for the analysis of the spiral structure.
In the case of the CO ratios, we only consider pixels that are detected above 3$\sigma$ 
in both input CO maps. 

\begin{table*}
\centering
\begin{minipage}{140mm}
\caption{Properties of the exponential disk in \cothree\/ and \cotwo\/ compared to the properties of the total gas exponential disk presented by \citet{hitschfeld09} and the exponential dust disk presented by \citet{meijerink05}.}
\begin{tabular}{lcc}
\hline
Exponential disk determined in: & Scale length & disk/total mass fraction \\
                                                   &  (kpc)            & \%\\
\hline
Molecular gas (\cothree\/)     &  6.9$\pm$0.4    & 14 \\
Molecular gas (\cotwo\/)       &  5.3$\pm$0.1    & 25 \\
Total gas (\cotwo\/ and \Hi)  & 7.4     & 61 \\
Dust                                        & 6.2    & 55 \\
\hline
\end{tabular}
\label{tab:scl}
\end{minipage}
\end{table*}

\subsection{Radial variation of CO, \Hi\/ and PAH emission}
\label{sec:radial-variation}

\begin{figure*}
\centering
\includegraphics[width=13.5cm]{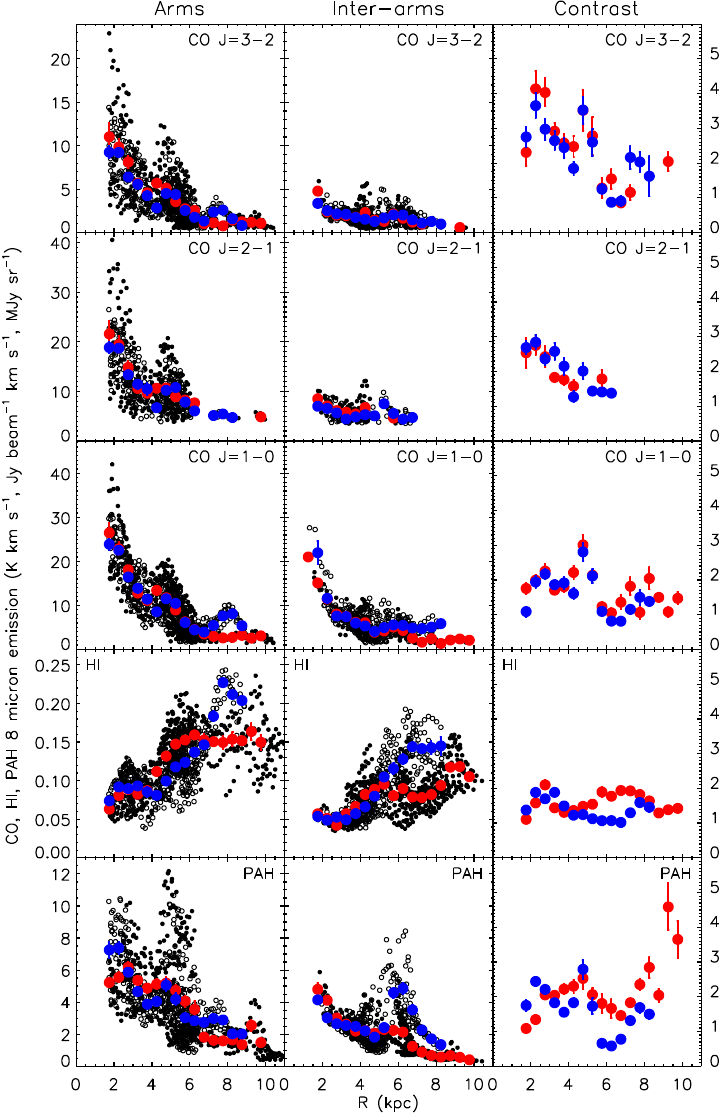}
\caption{Radial profiles of velocity-integrated CO \jthree, \jtwo, \jone\/  and \Hi\/ intensity and \pah\/ surface brighness in the two spiral arm and inter-arm regions (in units of K\,\kms\/ (\tmb) and Jy\,\kms, respectively). Arm and inter-arm regions are defined using a mask based on the \cothree\/ emission that is modified so that it also adequately describes the spiral structure observed in the other wavebands (see Section~\ref{sec:spiral} for further details); emission within the central 1.7\,kpc is not considered. Black points show individual 7.5\arcsec pixels with a signal-to-noise $>$3; filled points indicate Arm I and open points indicate Arm II (see text for definition). Red and blue points show the same data binned into 0.5\,kpc bins (approximately the size of the HARP-B beam-size), for Arm I and Arm II, respectively. Error bars show the error on the mean, and where not present are smaller than the symbol size.
The third column shows the arm--inter-arm contrast as a function of radius for the two spiral/inter-arm regions, i.e. the ratio of the first two columns; see Section~\ref{sec:arm-interarm-contrast} for definition).
}
\label{fig:spiral-profile}
\end{figure*}

Figure~\ref{fig:spiral-profile} shows the radial profiles of the 
\cothree, \cotwo, \coone\/ and \Hi\/ integrated intensities. The CO profiles of the two 
spiral arms are generally similar, with CO intensity decreasing with radius but with a 
strong `humps' at $\sim$5~kpc and $\sim$8~kpc. 
The \Hi\/ profiles also show some evidence of a broad 
humps at these radii. The former is the radius at which two strong regions of CO emission 
are seen in the spiral arms, to the north-east and the south-west of the centre.
Two notable differences are that the scatter appears to increase somewhat towards 
higher CO transitions, and there appears to be less scatter for the Arm II 
than Arm I. This is probably related to the interaction with the companion 
to the north of \mf. Conversely, there is remarkably little scatter for the inter-arm 
emission, with the shape of the profile broadly following that seen in the spiral arms 
(but without the broad hump). This is what may be expected if the inter-arm emission 
is diffuse emission underlying the disk. Despite these differences, the total profile 
(Figure~\ref{fig:spiral-profile}) for the whole galaxy disk shows a strong similarity 
between the shapes of the binned CO profiles on 0.5\,kpc scales. The emission scale lengths 
we measure for \cothree\/ and \cotwo\/ are shown in Table~\ref{tab:scl}. For comparison 
we also show the scale lengths of the total gas exponential disk 
presented by \citet{hitschfeld09} and the exponential dust disk presented by \citet{meijerink05}.

The \pah\/ surface density profile is seen to be similar to the profiles of \cothree\/ 
intensity but is very different to the profile of \Hi\/ emission, suggesting that 
in \mf\/ PAH emission is not associated with \Hi. 
While the radial profiles of \cothree\/ and \pah\/ emission are similar to one another, the 
radial profile of the \Hi\/ emission is clearly very different (Figure~\ref{fig:spiral-profile}), 
as is also obvious from simple inspection of the images in Figure~\ref{fig:maps-other}. 
As previously noted in Section~\ref{sec:images:otherwave}, it is also clear from Figure~\ref{fig:spiral-profile} 
that the intensities of CO and \Hi\/ are anti-correlated, i.e. that the 
two species are complementary, which suggests that the ratio \Hi/\Htwo\/ is a strong, and 
apparently continuous, function of radius.

\begin{figure*}
\centering
\includegraphics[width=17cm]{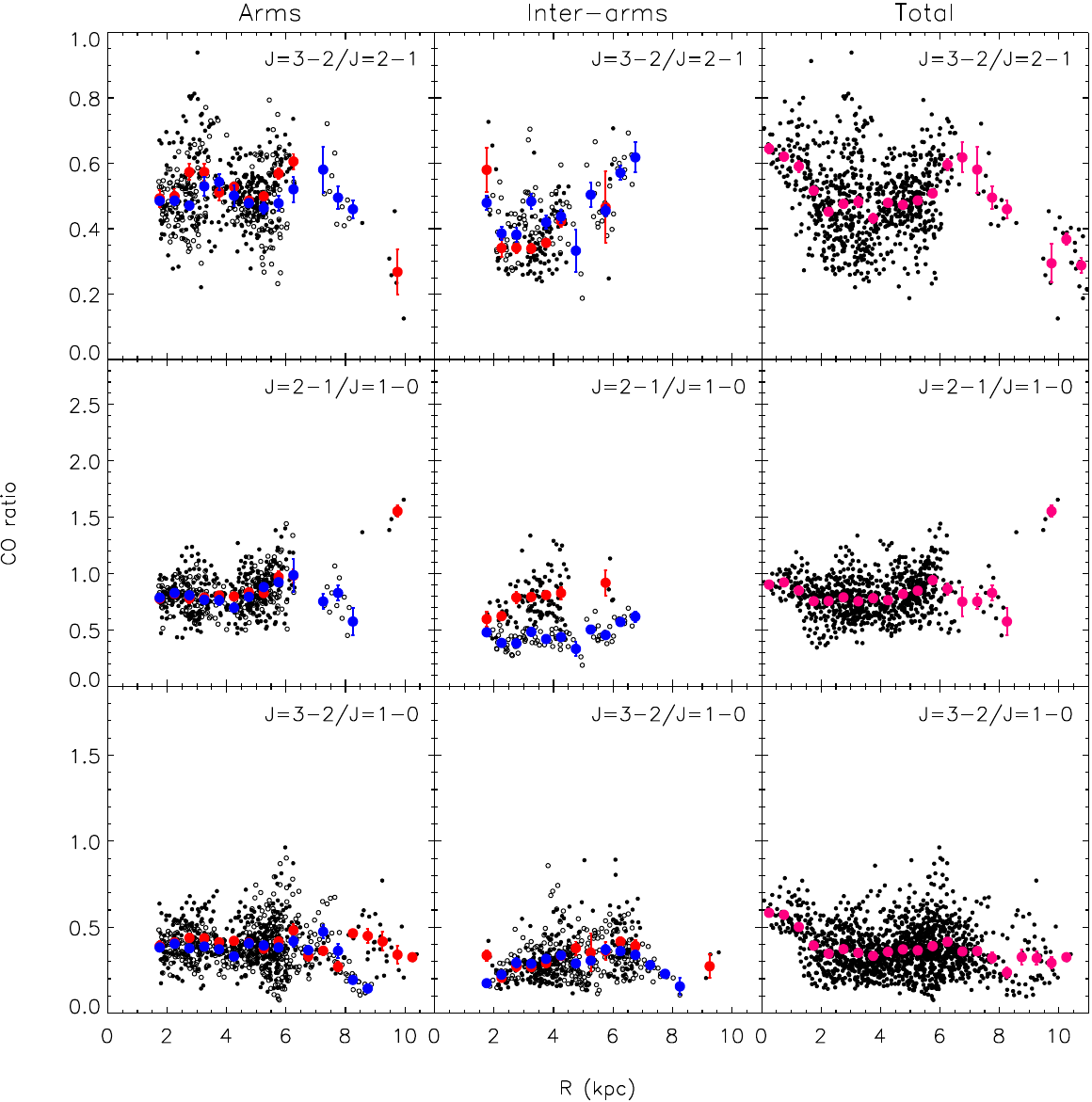}
\caption{CO line ratios as a function of distance along the spiral arms and along the inter-arm 
regions, starting from a radius of 1.7\,kpc. Symbols are the same as for Figure~\ref{fig:spiral-profile}. 
Median line ratios are listed in Table~\ref{tab:coratios}. 
The final column shows the total radial profile for the galaxy, including the central 1.7\,kpc. 
CO emission is detected everywhere, in both the relevant transitions, out to a radius of $\sim$4\,kpc. Thus, out to 
$\sim$4\,kpc the binned profiles shown here are equivalent to the profile that would be obtained by 
averaging in a series of concentric annuli.
}
\label{fig:coratio-profile}
\end{figure*}

\subsection{Radial variation of arm--inter-arm contrast}
\label{sec:arm-interarm-contrast}

In the right-hand panel of Figure~\ref{fig:spiral-profile} we show the contrast between arm and 
inter-arm emission as a 
function of radius for the two spiral arms and inter-arm regions. 
Contrast is 
defined as the ratio of the mean spiral arm velocity-integrated intensity versus the mean 
inter-arm integrated intensity in bins of 0.5~kpc (i.e. the ratio of the red points shown in 
Figure~\ref{fig:spiral-profile}). 
The average arm/inter-arm contrast values for \cothree,  \cotwo, \coone\/ and \Hi\/ emission, 
as well as \pah\/ surface density, are given in Table~\ref{tab:contrast}. As expected from 
simple inspection of the images in Figures~\ref{fig:maps-ico32}--\ref{fig:maps-otherco} 
there appears to be a trend for 
arm-interarm contrast to increase towards higher CO transitions. 
For \cotwo, a value of 2.9 was previously found by \citet[][]{hitschfeld09} 
who used used the same dataset as the present work but with different method of defining the arm and inter-arm 
regions, while we find a value of 2.1. Both values are in reasonable agreement with \citet[][]{garcia-burilloetal93b}.

It is clear that for \Hi\/ and \coone\/ emission 
there is little variation in the arm-inter-arm contrast with radius over $\sim$10~kpc. 
Conversely, for the higher CO transitions the contrast is high in the central few 
kiloparsecs and clearly 
decreases towards larger radii. The trend is particularly clear for \cothree\/ emission, 
which decreases from a contrast of $\sim$ 4 at 2~kpc to $\sim$1 at $\sim$6.5~kpc 
(the co-rotation radius).

\begin{figure*}
\centering
\includegraphics[width=17cm]{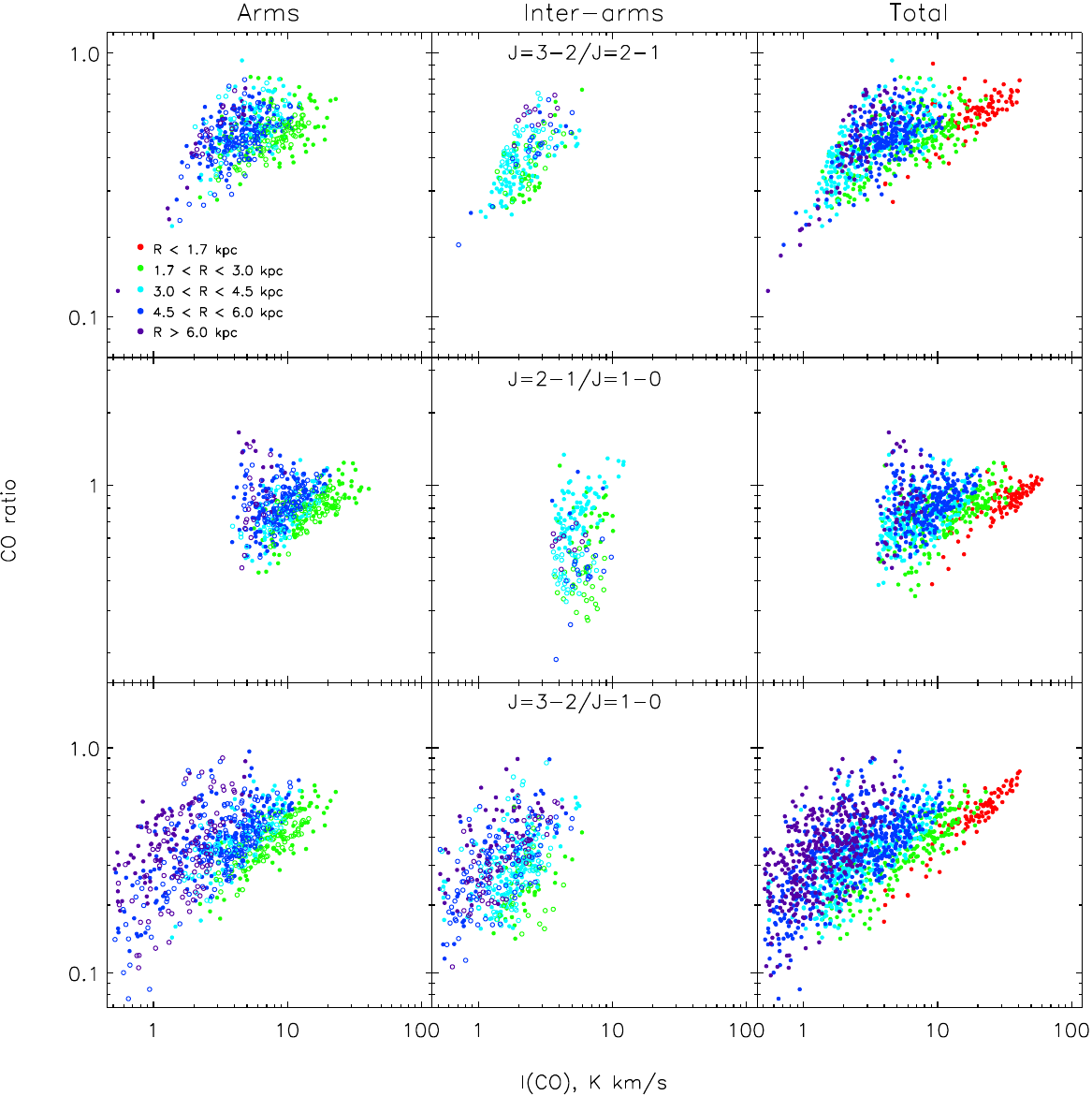}
\caption{CO $J$=3-2/$J$=2-1, $J$=2-1/$J$=1-0, and $J$=3-2/$J$=1-0 line intensity ratios (rows) 
plotted as a function of \cothree, \cotwo, and \cothree\/ line intensity, respectively, for the spiral arms (left panels), inter-arm regions (centre panels) and entire galaxy disk (right-hand panels). Only pixels with 
emission detected above 3$\sigma$ in both CO transitions are shown. In the right-hand column we 
colour-code the plotted points according to galactocentric radius as indicated in the top right panel: 
$<$1.7\,kpc (red), 1.7--3\,kpc (green), 3--4.5\,kpc (cyan), 4.5--6\,kpc (blue), and $>$6\,kpc (purple). 
For the left and centre panels, filled and open points show Arm I and II, respectively.}
\label{fig:coratio-ico}
\end{figure*}

\subsection{Radial variation of CO line ratios}

\begin{table}
\caption{Average arm-interarm contrast.}
\begin{tabular}{lc}
\hline
Transition & Arm-interarm contrast\\
\hline
\cothree\/     & 2.4$\pm$1.1  \\
\cotwo\/       & 2.1$\pm$0.5  \\
\coone\/      & 1.7$\pm$0.5 \\
\Hi\/            & 1.6$\pm$0.3 \\
\pah\/         & 2.2$\pm$0.8 \\
\hline
\end{tabular}
\label{tab:contrast}
\end{table}

In Figure~\ref{fig:coratio-profile} we show the CO ratios \Rthtw, \Rthon, and \Rtwon\/ 
as a function of distance along the spiral arms and inter-arm regions. 
The last column shows the radial profile for the entire galaxy.
Despite considerable scatter in the individual pixel values, averaged on 0.5~kpc scales 
the variation in CO ratio in the spiral arms is surprisingly small, typically much less 
than a factor of two, both within an individual spiral arm and compared between the two 
spiral arms. The main notable difference between Arm I and Arm II occurs for \Rthon\/ at 
$\sim$7~kpc, near the co-rotation radius. Here, the value of \Rthon\/ for Arm I decreases by 
a factor of about two while for Arm II the ratio increases by a similar factor. While the 
values for the other CO ratios also increase for Arm II at this radius, for Arm I virtually 
no significant \cotwo\/ emission is detected beyond $\sim$6~kpc.
The CO ratio 
profiles for the total galaxy disk again emphasize that the variation on 0.5~kpc 
scales appears to be within a factor of two, whether in individual spiral structures 
or averaged over the whole galaxy. It is interesting to note, however, that 
while the total profile is almost flat from 2--4~kpc in all three transitions, 
in the individual spiral arms there is nonetheless some variation -- in the case of 
\Rthtw\/ as much as a factor of two. Comparing Arm~I and Arm~II it is 
clear why this is -- in the eastern arm the ratio is decreasing while in the western 
arm it is increasing. Thus, this shows that in only taking radial averages over the 
whole galaxy one would `miss' a factor of two variation between the two spiral arms.

Median line ratios are \Rthtw\/$\sim$0.5 in the spiral arms and 
$\sim$0.4 in the inter-arm regions. 
For \Rtwon, average values are $\sim$0.4 in the spiral arms and 
$\sim$0.3 in the inter-arm regions. 
We note that we find lower median values of \Rtwon\/ and \Rthtw\/ 
for inter-arm I than for inter-arm II. 
While this may hint at a small systematic difference in the average CO ratio between the arm 
and inter-arm regions we note that the difference is only of the order of one standard deviation 
(Section~\ref{sec:coratio}).
Lower line ratios suggests that the molecular gas may on average be cooler and/or less dense,
so it is not unexpected to find higher values in the spiral arms which are clearly the sites of 
star formation (see e.g. Figures~\ref{fig:maps-polar-co} and~\ref{fig:maps-polar-other}) 
and thus should host warmer, denser, gas. 
Our values of \Rthtw\/ and \Rtwon\/ are somewhat lower than found in 
previous works -- \citet{garcia-burilloetal93a} and \citet{kramer05} find values of $\sim$0.7 
over the whole disk and in the central region, respectively.

\begin{table}
\caption{Parameters of linear fits in Figures~\ref{fig:coratio-ico-fit} and~\ref{fig:co32-pah-fit}.}
\centering
\begin{tabular}{cccc}
\hline
Radius & A & B & $r_{p}$\\
(kpc) &  &  & \\
\hline
\multicolumn{4}{c}{Relationship between CO~\jthree/\jone\/ ratio and \cothree\/ intensity:}\\
R$<$1.7 & 0.568(0.024) & -1.019(0.032) & 0.89 \\
1.7$<$R$<$2.7 & 0.515(0.020) & -0.878(0.019) & 0.89 \\
2.7$<$R$<$3.7 & 0.549(0.019) & -0.761(0.011) & 0.80 \\
3.7$<$R$<$4.7 & 0.555(0.038) & -0.683(0.016) & 0.58 \\
4.7$<$R$<$5.7 & 0.551(0.022) & -0.678(0.010) & 0.68 \\
5.7$<$R$<$6.7 & 0.791(0.042) & -0.630(0.013) & 0.69 \\
R$>$6.7 & 1.104(0.104) & -0.591(0.017) & 0.46 \\
\multicolumn{4}{c}{Relationship between \pah\/ and \cothree\/ intensity:}\\
R$<$1.7 & 0.869(0.044) & 0.016(0.059) & 0.85 \\
1.7$<$R$<$2.7 & 0.556(0.024) & 0.285(0.019) & 0.80 \\
2.7$<$R$<$3.7 & 0.643(0.017) & 0.230(0.011) & 0.91 \\
3.7$<$R$<$4.7 & 0.711(0.019) & 0.202(0.010) & 0.89 \\
4.7$<$R$<$5.7 & 0.805(0.021) & 0.209(0.012) & 0.88 \\
5.7$<$R$<$6.7 & 0.895(0.031) & 0.249(0.013) & 0.77 \\
R$>$6.7 & 0.989(0.049) & 0.160(0.015) & 0.45 \\
All  & 0.718(0.09) & 0.220(0.005) & 0.89\\
\hline
\end{tabular}\\
Column 1: radial intervals (in kpc); 
Columns 2 and 3: parameters of linear fits to log(\Rthon) versus log(\cothree\/ intensity) and log(\pah\/ surface brightness) versus log(\cothree\/ intensity); uncertainties are given in parentheses; 
Column 4: Pearson correlation coefficient.
\label{tab:fits}
\end{table}

However, as can be seen in the 
last panel of Figure~\ref{fig:coratio-profile}, our value of \Rthtw\/ in the central 
kiloparsec is $\sim$0.6, with values for individual lines of sight in the inner spiral arms 
ranging up to $\sim$0.8, 
while our value for \Rtwon\/ are $\sim$0.5--0.6 in the central kiloparsec. Thus, within the uncertainties, 
we find our values of \Rthtw\/ to be in agreement with those in the literature.
Our values of \Rthon\/ are also in good agreement with those found in previous works. 
We find median values of $\sim$0.4 in the spiral arms, 
$\sim$0.2 in the inter-arms, and $\sim$0.54 in the central kiloparsec. 
For comparison, \citet{garcia-burilloetal93a} found values of $\sim$0.6 in the spiral arms within 
the central 1\arcmin.

For each of the CO line ratios, we generally find no significant difference between the mean values 
in Arms I and II, nor do we find any significant difference between the two inter-arm regions.
Thus, while there is clearly variation between the individual spiral arms at a given radius, 
as described above, when averaged over distance the CO ratios appear to be similar.

The ratio \Rtwon\/ for the spiral arms shows some evidence of an increasing trend towards 
larger galactocentric radii, while there is no clear trend for \Rthtw\/ or \Rthon. 
The lack of a clear trend for \Rthon\/ to vary with radius when studied over either all regions of 
emission, (the right-hand panel of Figure~\ref{fig:coratio-profile}) or just the spiral arms 
(the left-hand panel of Figure~\ref{fig:coratio-profile}), 
is in agreement with the findings for NGC~2403 by \citet[][]{bendo10}. 
However, for \mf, in the inter-arm regions there appears to be a trend for all the CO ratios to 
increase with radius.

\subsection{CO line ratios and the $X$ factor}
\label{sec:coratio-x}

The large pixel-to-pixel variation in the CO ratio \Rthon\/ described above, together with the 
suggested different behaviour with radius for the spiral arms and inter-arm regions, implies that 
obtaining molecular gas column densities from the \cothree\/ data by applying a simple scaling 
term to match it to the $X$-factor determined for \coone\/ is likely to introduce large 
uncertainties for sub-kiloparsec scales. As in \citet[][]{bendo10}, we investigate this further 
by looking at the relationship between CO line ratio and CO surface brightness. We plot this 
for \Rthtw\/, \Rtwon\/ and \Rthon\/ in Figure~\ref{fig:coratio-ico}. In the first two columns we 
plot separately the regions of spiral arm and inter-arm emission, as described for the previous 
figures, and in the last column we show the entire galaxy disk (where emission is detected at 
$>$3$\sigma$ in both CO lines), colour coded for galactocentric radius. We can immediately note 
several features of the distribution of the CO ratio with CO surface brightness that was not 
apparent for the relationship with galactocentric radius (Figure~\ref{fig:coratio-profile}).
First, in the inter-arm regions, for a given CO line intensity the CO ratios \Rthtw\/ 
and \Rtwon\/ tend to be higher in Arm II than in Arm I. This difference is not apparent for 
the inter-arm in \Rthon\/, and is also not obvious for the spiral arms. Second, for \Rthtw\/ 
and \Rthon\/ the slope of the relationship between CO ratio and CO surface brightness appears to 
be steeper for the inter-arm regions than for the spiral arms. Thirdly, for all three CO ratios 
we see that the relationship with CO surface density clearly varies with galactocentric radius, 
and that the dependence of the relationship on radius appears to be greater for \Rtwon\/ and 
\Rthon\/ than for \Rthtw. 
At the centre of the galaxy, where the highest CO surface densities are located, there is little 
variation in the CO ratio for a given CO line intensity. However, as galactocentric 
distance increases, and CO surface densities decrease, there is increasingly large variation 
in the CO ratio for a given \ico. 

\begin{table}
\caption{Parameters of relationship in Equation~\ref{eq:rel}.}
\centering
\begin{tabular}{lrrr}
\hline
Radius & A & B & C \\
(kpc) & & & \\
\hline
R$<$1.7 & 0.6 & 0.19 & 1.3 \\
1.7$<$R$<$2.7 & 0.6 & 0.12 & 1.2 \\
2.7$<$R$<$3.7 & 0.6 & 0.12 & 1.2 \\
3.7$<$R$<$4.7 & 0.6 & 0.12 & 1.2 \\
4.7$<$R$<$5.7 & 0.6 & 0.12 & 1.3 \\
5.7$<$R$<$6.7 & 0.9 & 0.10 & 1.3 \\
R$>$6.7 & 1.2 & 0.09 & 1.3 \\
\hline
\end{tabular}\\
Column 1: radial intervals (in kpc). The mean value in each interval is used; this is 
the central value in each radial interval, except for R$<$1.7\,kpc where the mean radius is 1.1\,kpc and 
R$>$6.7\,kpc where the mean radius os 7.6\,kpc; 
Columns 2, 3 and 4: parameters of the linear fits to Equation~\ref{eq:rel}). 
\label{tab:rel}
\end{table}

In Section~\ref{sec:coratio} we noted an apparent trend for \Rthon\/ in the inter-arm regions to increase with 
radius. From inspection of Figure~\ref{fig:coratio-ico} (lower middle panel) we can now see clearer evidence of this. 
Here, for a given CO intensity high values of the ratio always occur at higher radii. This is in contrast to 
the behaviour seen in the spiral arms (lower left panel of~Figure~\ref{fig:coratio-ico}), 
where larger radii exhibit higher ratios for lower 
CO intensities than small radii, and likewise smaller radii exhibit low ratios for higher CO intensities than 
large radii. These trends can be explained as follows. The general dropoff in all CO ratios with radius, for both 
spiral arms and inter-arm regions, is likely explained by the decreasing beam-averaged molecular gas column density 
with radius. For the spiral arms, which contain star-forming regions, 
high ratio values reflect high gas densities and/or temperatures. The trend with radius suggests that active 
star-formation sites occur at lower CO intensities at larger radii, which may reflect an underlying star formation 
threshold \citep[e.g. the Toomre Q parameter;][]{toomre72}. One would not expect the same trend in 
the inter-arm regions since they contain little or no active star formation (see Figure~\ref{fig:spiral}, which 
shows that with the exception of inter-arm II to the north-east at large radii there are no significant regions of 
star formation that are not included within our spiral arm mask). There are a number of possible explanations for 
the apparent trend with radius in the inter-arm regions. One possibility is that the temperature and/or density 
is increasing with radius, but this does not seem likely in the inter-arm regions. Another possibility is 
that the inter-arm medium consists of a diffuse molecular medium (dominating the lowest emission lines most) 
in which dense clouds are embedded. Loss of the diffuse medium at large radii would result in the observed behaviour. 
An alternative possibility is that the CO begins to become optically thin at large radii. Under typical excitation conditions 
all low CO lines are optically thick, and if the density and temperature are high enough \cothree\/ will have 
higher optial depth than the lower lines. Thus, if cloud column densities in the inter-arm region decrease with radius,  
\coone\/ may become optically thin while \cothree\/ remains optically thick, hence producing the 
observevd trend. 

Variation of the CO ratio with CO surface density suggests a relation to star formation and local 
heating. Variation with radius, on the other hand, suggests a relation to a more diffuse component 
in the disk. Thus, Figure~\ref{fig:coratio-ico} clearly suggests that the \coone\/ line emission 
is also tracing a more diffuse and radius-dependent component of molecular gas than the higher CO 
transitions.

While it is clear that the variation of CO ratio with CO 
surface density is dependent on galactocentric distance, the important result from this plot is 
that a simple relationship between the CO ratio \Rthon\/ and the \cothree\/ integrated intensity 
cannot be used to scale the \coone-derived $X$-factor for use in determining molecular gas 
surface density from the \cothree\/ line without talking into account the spatial location of the 
emission in the galaxy -- the variation in the relationship implies that applying a global scaling 
for sub-kiloparsec scales will overestimate the molecular gas surface density at large 
galactocentric radii and underestimate it at small radii.

\begin{figure}
\centering
\includegraphics[width=8cm]{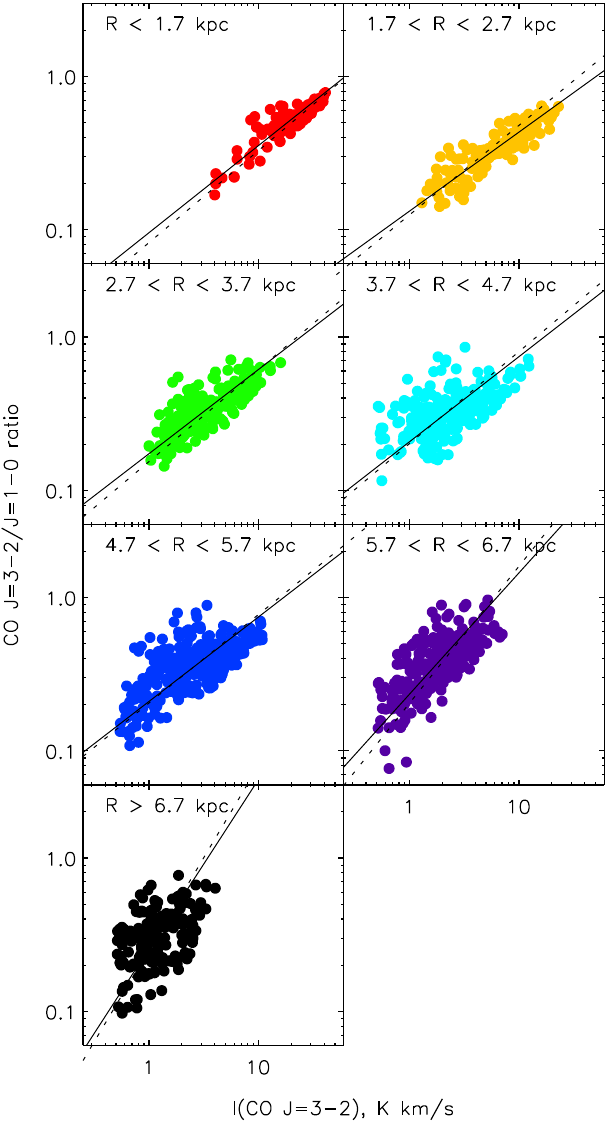}
\caption{Relationship between CO~\jthree/\jone\/ line intensity ratios and \cothree\/ intensity at
different galactocentric radii (over the entire galaxy disk where individual pixels are detected above 3$\sigma$ in both CO transitions). 
We colour-code the plotted points according to galactocentric radius at smaller radial intervals than 
for Figure~\ref{fig:coratio-ico}; here we show the central emission with R$<$1.7\,kpc and 1\, kpc 
intervals thereafter. The solid lines show bi-variate least squares fits, whose parameters are 
listed in Table~\ref{tab:fits}. The dashed lines show the radius-dependent relationship given by 
Equation~\ref{eq:rel} using the mean distance of pixels in each radial bin.
}
\label{fig:coratio-ico-fit}
\end{figure}

\begin{figure}
\centering
\includegraphics[width=8cm]{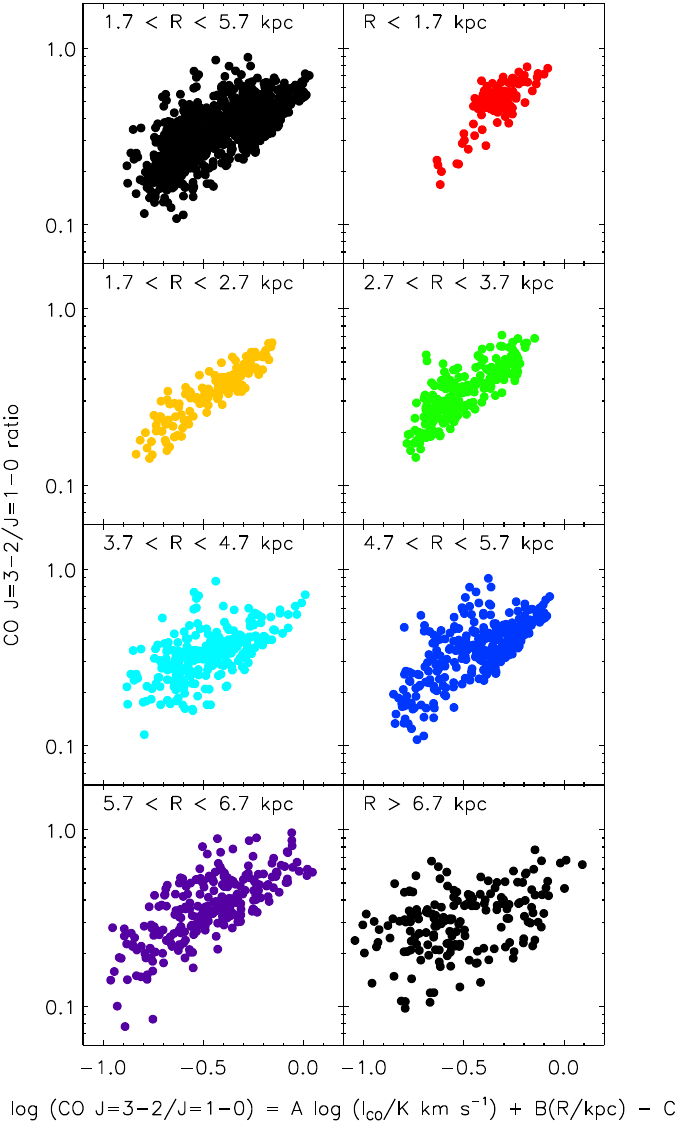}
\caption{Relationship between CO~\jthree/\jone\/ intensity ratio and the relationship given by the 
right-hand side of Equation~\ref{eq:rel}, for all pixels 
detected above 3$\sigma$.
Points are colour-coded as in Figure~\ref{fig:coratio-ico-fit}.
}
\label{fig:coratio-ico-rel}
\end{figure}

In Figure~\ref{fig:coratio-ico-fit} we investigate the radial dependence of the relationship 
between CO ratio and \cothree\/ intensity, by plotting the data at radial intervals of 
$\sim$1\,kpc and fitting a bi-variate linear function for each bin (shown as solid lines in 
Figure~\ref{fig:coratio-ico-fit}). The parameters of the linear fits are given in Table~\ref{tab:fits}. 
We find that the Pearson's correlation coefficient decreases considerably with galactocentric 
radius, from 0.95 in the centre to 0.5 at $\sim$7\,kpc.

The dashed lines in Figure~\ref{fig:coratio-ico-fit} show radius dependent relations given by a linear function of the CO ratio 
with both \cothree\/ intensity and galactocentric radius such that
\begin{equation}\label{eq:rel}
\textrm{log}\left( \textrm{CO}~J=3-2/J=1-0 \right) = A~\textrm{log} \left( \frac{I_{CO}}{\textrm{K\,\kms}} \right) + B \left( \frac{R}{\textrm{kpc}} \right) - C
\end{equation}
for the parameters given in Table~\ref{tab:rel}. 
We find that a relationship with a slope of 0.6 reasonably 
describes the emission between radii of 1.7 and 5.7~kpc, but that different relationships are 
needed to describe both the central regions (shallower slope) and regions at radii larger than $\sim$6~kpc (steeper slope). 
Nonetheless, a global fit across all radii yields a slope of 0.6, which is comparable to the value found by \citet[][]{bendo10} for NGC~2403. 
The relationship between the CO ratio and Equation~\ref{eq:rel} for individual pixels is shown in 
Figure~\ref{fig:coratio-ico-rel}. The scatter in the relationship clearly increases with radius, 
as indicated by the decreasing correlation coefficient (Table~\ref{tab:fits}).

\subsection{Relations between \cothree, PAH and 24\micron\/ emission as a function of the spiral structure}
\label{sec:copah}

It is interesting to compare the relationships between PAHs, cold dust, molecular gas and atomic gas emission 
in \mf\/ with those in other late-type galaxies, 
and in particular whether PAH emission is a tracer of gas and dust in the cool ISM. 
A good target for a simple comparison is the nearby 
\citep[$\sim$\,3~Mpc;][]{freedman01} SABcd spiral galaxy NGC\,2403, whose relationships between 
these properties were studied by \citet{bendo10}.
While previous authors have found that PAH emission in nearby galaxies is strongly correlated 
with cold dust emission \citep[e.g.][]{bendo06,zhuetal08}, and some authors \citep[e.g.][]{reganetal06} 
have found a correlation between PAH emission and CO line emission in nearby galaxies, \citet{bendo10} find that, 
in NGC\,2403,  the \cothree\/ and PAH surface brightnesses are uncorrelated on sub-kpc scales. 

For both \mf\/ and NGC\,2403, as for many late-type spiral galaxies, the strongest regions of star formation are located 
outside the centre of the galaxy. 
Qualitatively, the structures seen in the \cothree, PAH, 24\micron\/ and \Ha\/ images for \mf\/ and NGC\,2403 
are similar. Exceptions for \mf\/ are the inner part of Arm II, the south-west where CO emission is seen to 
trace the spiral arm but relatively little PAH emission is seen, and the very centre where there is a dip 
in the PAH emission but not CO (the latter difference, however, may be due to the different spatial resolutions). 
Conversely, an exception for NGC\,2403 is a hole in the CO emission at the galaxy centre that is not seen in 
PAH emission \citep{bendo10}.

\begin{figure*}
\centering
\includegraphics[width=17cm, clip]{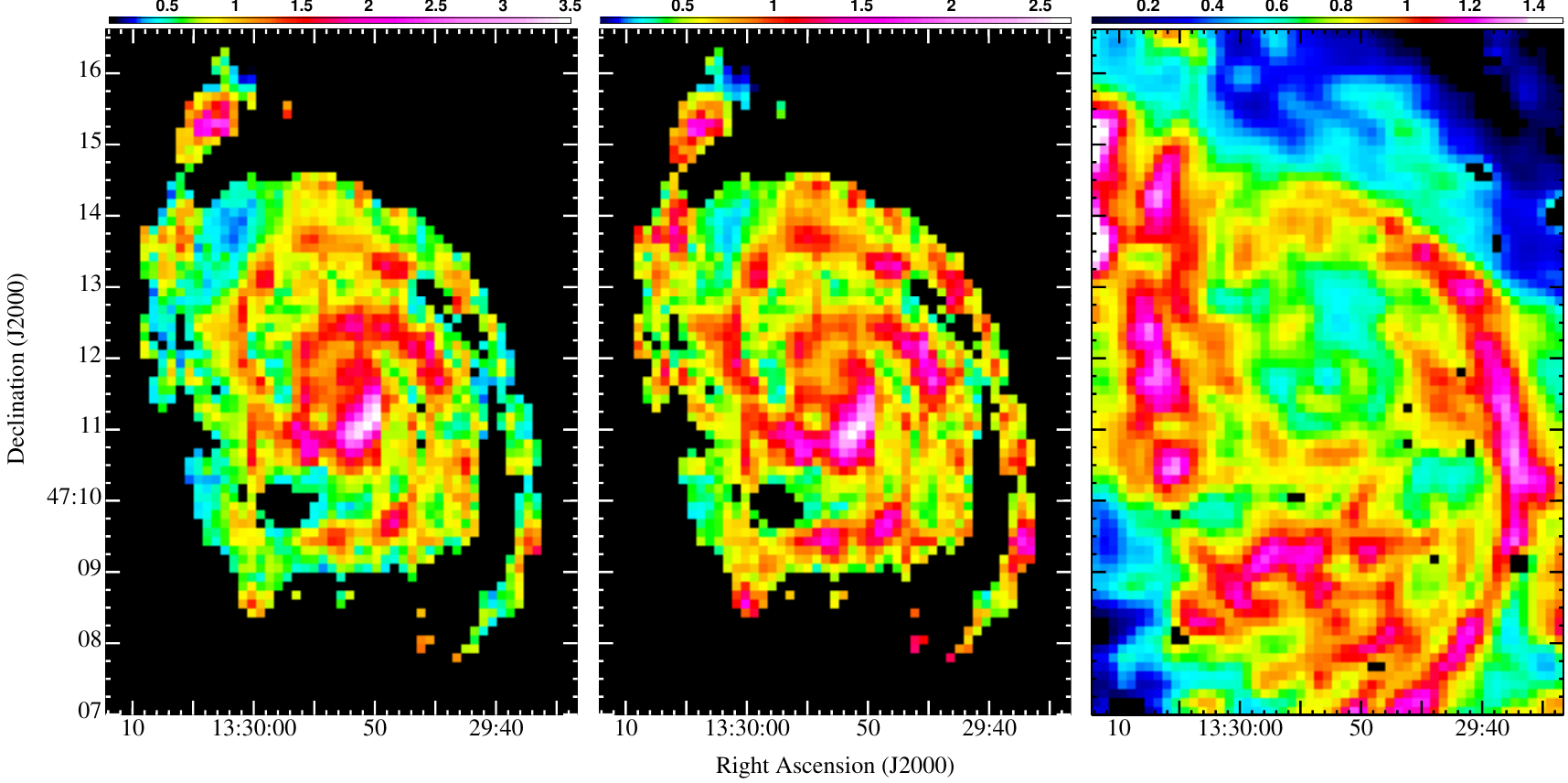}
\caption{Ratio maps. 
Left: \cothree\/ integrated intensity (\kkms) to \pah\/ surface 
brightness (MJy\,sr$^{-1}$). 
Centre: \cothree\/ integrated intensity (\kkms) to 24\micron\/ surface 
brightness (MJy\,sr$^{-1}$). 
Right: \pah\/ to 24\,\micron\/ surface density (both in units of MJy\,sr$^{-1}$). 
The two CO ratio maps are shown on the same colour scale.
All maps were produced as described in Sections~\ref{sec:ancillary} 
and~\ref{sec:copah}.}
\label{fig:maps-ratio}
\end{figure*}

\begin{figure*}
\centering
\includegraphics[width=17cm]{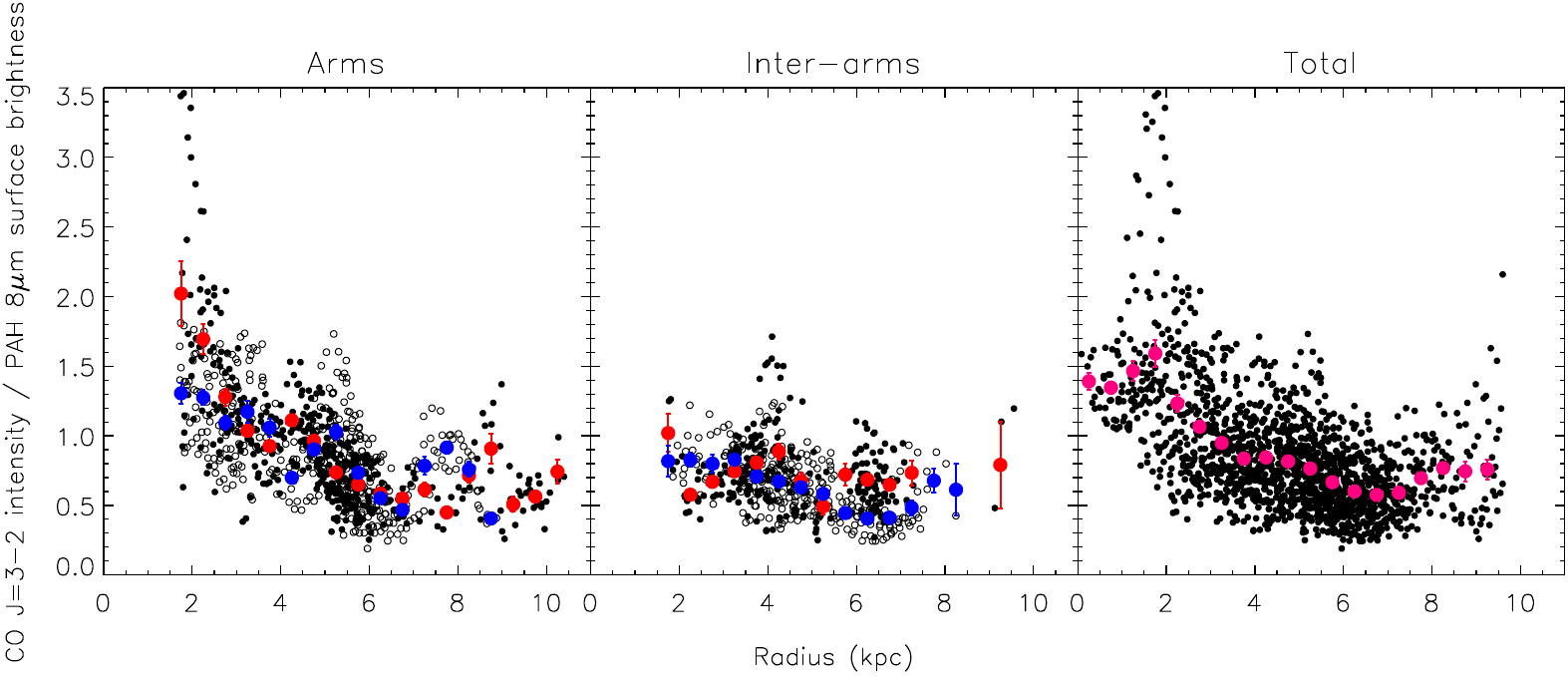}\\
\includegraphics[width=17cm]{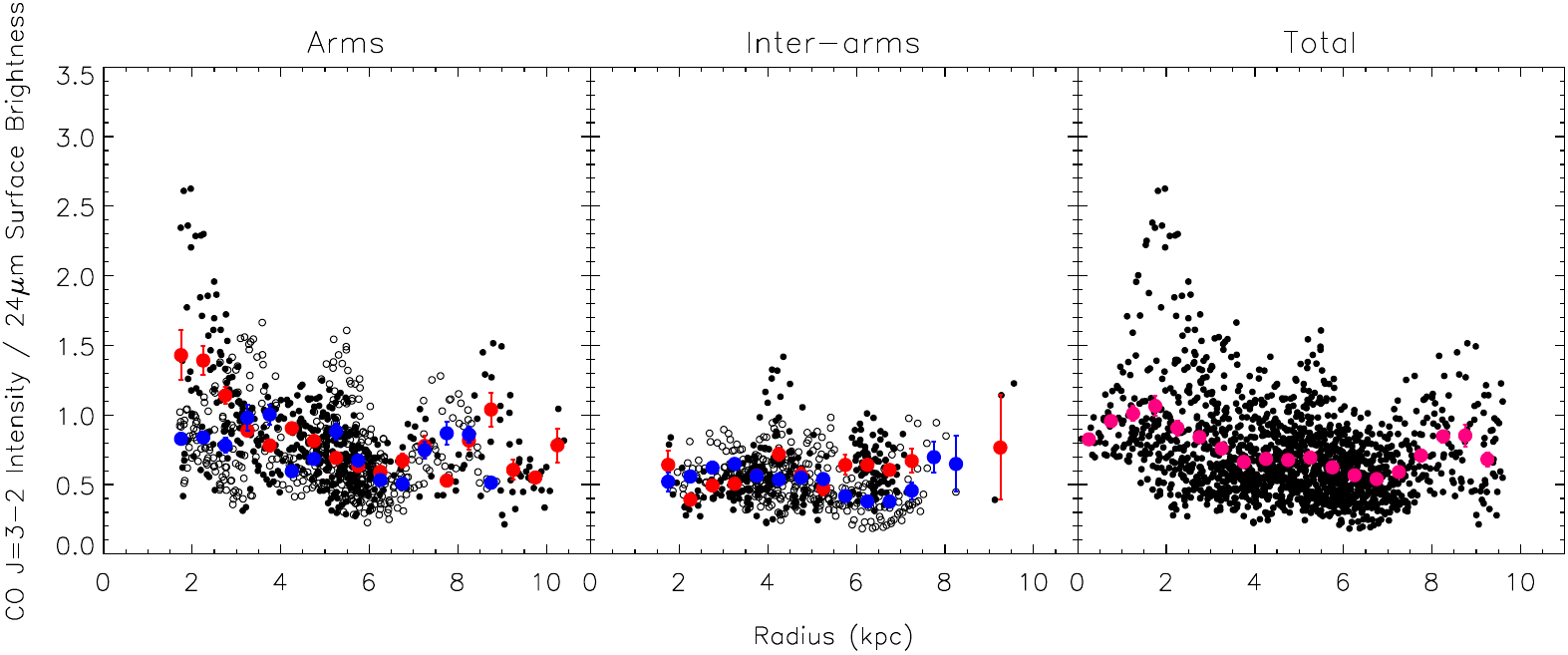}\\
\includegraphics[width=17cm]{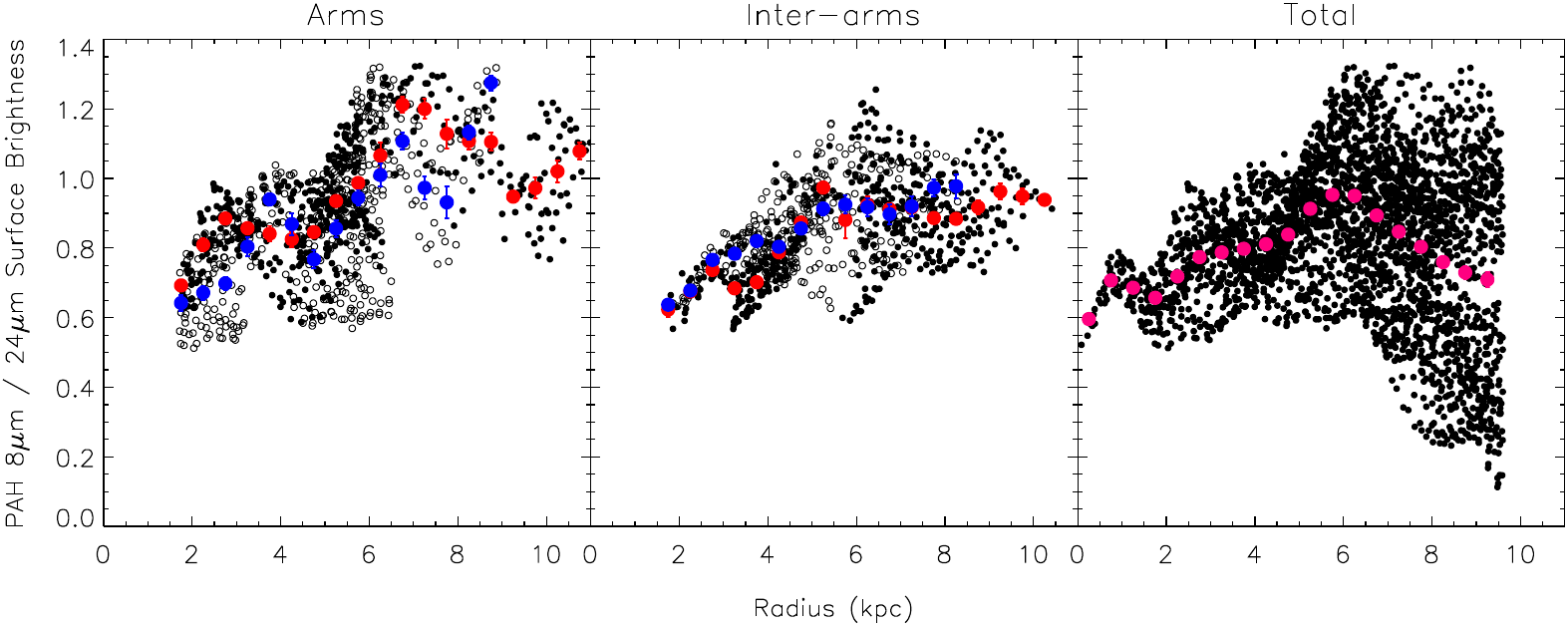}
\caption{Ratios as a function of distance along the spiral arms and inter-arm 
regions, starting from a radius of 1.7\,kpc. 
Top: \cothree\/ intensity (in \kkms) to \pah\/ surface brightness (in MJy\,sr$^{-1}$. 
Middle: \cothree\/ intensity to 24\,\micron\/ surface brightness.
Bottom: \pah\/ to 24\,\micron\/ surface density. 
Symbols are the same as for 
Figures~\ref{fig:spiral-profile} and~\ref{fig:coratio-profile}.
}
\label{fig:profiles-ratio}
\end{figure*}

\begin{figure*}
\centering
\includegraphics[width=16cm]{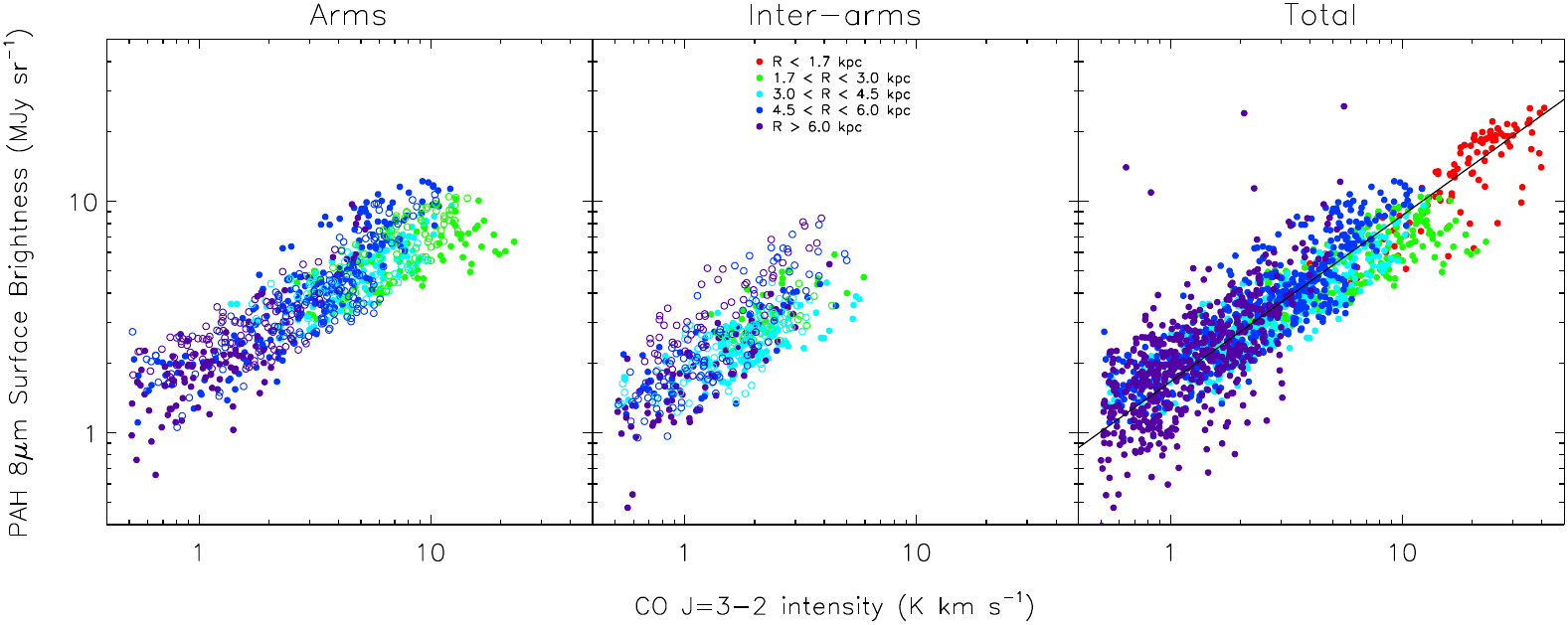}\\
\includegraphics[width=16cm]{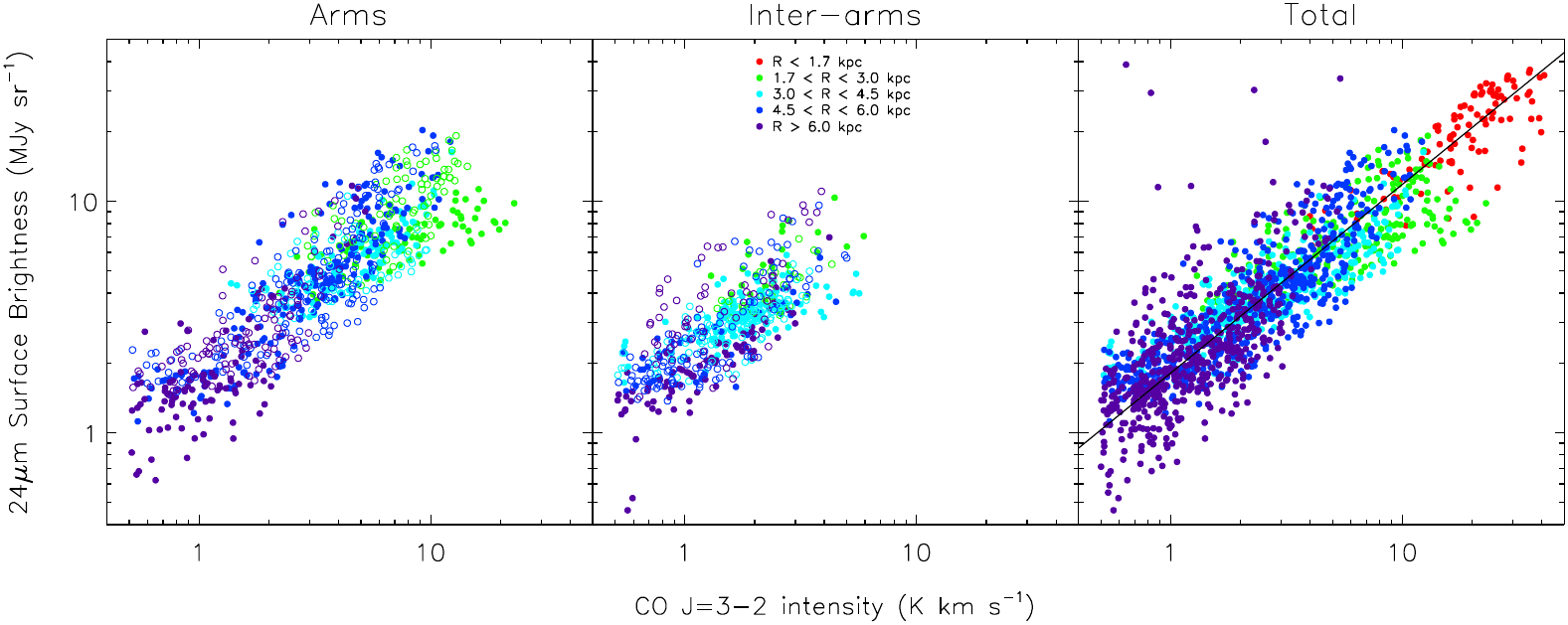}
\caption{Relationships between \cothree\/ intensity and \pah\/ surface brightness (top) 
and \cothree\/ intensity and 24\micron\/ surface brightness (bottom)
for different locations in \mf: spiral arms, inter-arm regions, and the entire galaxy including 
the outer disk (from left), 
for all pixels detected above 3$\sigma$. 
Symbols are the same as for Figure~\ref{fig:coratio-ico}.
For emission over the entire galaxy (right-hand panels),  both relationships have a Pearson 
correlation coefficent of 0.89. The solid lines show a bi-variate least squares fit given by 
log (\pah) = (0.718$\pm$0.009) log (\cothree) + (0.220$\pm$0.005) and 
log (24\,\micron) = (0.814$\pm$0.010) log (\cothree) + (0.257$\pm$0.006).
}
\label{fig:co32-pah-spiral}
\end{figure*}

\subsubsection{Relationships between \cothree\/ and \pah\/ emission}

Following \citet[][]{bendo10} we used the \cothree\/ integrated intensity and \pah\/ surface brightness images presented in 
Figures~\ref{fig:maps-ico32} and~\ref{fig:maps-other} to produce a map and radial profile of the 
\cothree\//\pah\/ emission ratio (left-hand panel of Figure~\ref{fig:maps-ratio} and the top panel of Figure~\ref{fig:profiles-ratio}, respectively). 
The two input images are in units of \kkms\/ and MJy\,sr$^{-1}$, 
respectively. As before, 
only pixels detected above 3$\sigma$ in the input images are shown. 

The \cothree/\pah\/ ratio ranges from 0.04 to 3.5 across \mf\/, with a median value of 0.8 and 
standard deviation of 14.1. 

The \cothree/\pah\/ ratio appears to broadly follow the spiral structure, with the highest 
values in the spiral arms and the lowest values in the inter-arm regions. 

In the spiral arms we find a mean \cothree/\pah\/ ratio of unity, and values of $\sim$0.7 in 
the inter-arm regions. The ratio appears to decrease with radius for the spiral arms, falling 
by a factor of about two between $\sim$2 and 6~kpc. 
There appears to be no obvious overall 
trend with radius for the inter-arm regions, but we note, however, that the presence of strong features in the 
inter-arm profile at $\sim$4 and $\sim$6~kpc, which are similar to features in the spiral arms, 
suggests that the spiral arms traced in \pah\/ emission may be broader than for CO emission. 
It is clear that the variation of the ratio of \cothree\/ to \pah\/ emission in 
\mf\/ is very different to the profile that was found for NGC\,2403 by \citet[][]{bendo10}, 
who showed that in that galaxy the ratio increases with radius but decreases sharply in the 
central region; NGC\,2403 typically has lower ratio values than found for 
\mf. It seems likely that the \cothree\/ and \pah\/ properties 
for \mf\/ are very different from those for NGC\,2403. While \citet[][]{bendo10} showed that, 
for NGC\,2403, the \cothree\/ and \pah\/ emission does not trace the same structures on small spatial scales, 
in the present work for \mf\/ we find that the \cothree\/ and \pah\/ emission {\it do} 
trace the same structures on sub-kiloparsec scales, as is also clear from simple visual inspection 
of Figures~\ref{fig:maps-other}--\ref{fig:maps-polar-other}.

The most striking feature is that the highest values of the \cothree/\pah\/ ratio are located 
$\sim$1.5--2~kpc to the south-west, in Arm I (see Figure~\ref{fig:maps-ratio}). Here, ratio values are up to a factor of 
two higher than typically seen elsewhere in the spiral arms. 
From inspection of Figure~\ref{fig:maps-other}, it is 
evident that this region of enhanced \cothree-to-\pah\/ emission corresponds to a region 
of reduced \pah\/ and 24\micron\/ surface brightness. Here, there are no strong 
features apparent in the \Ha\/ image, but an integral field spectroscopy study of nebular 
emission lines in the central regions 
by \citet{blanc09} does show an \Ha\/ emission feature at this location after extinction 
correction and subtraction of a contribution from diffuse ionised gas. 
This is also just south along the spiral arm from the location where 
we have noted the highest \cothree\/ and \cotwo\/ arm/inter-arm contrast 
(Section~\ref{sec:images}).
Conversely, two obvious regions of low \cothree/\pah\/ ratio occur at symmetrically 
opposite locations $\sim$6~kpc to the north-east (Arm I) and south-west (Arm II) that 
are coincident with the locations of bright peaks in the \pah, 24\micron\/ and \Ha\/ images 
that are not peaks in CO. 

\begin{figure}
\centering
\includegraphics[width=8cm]{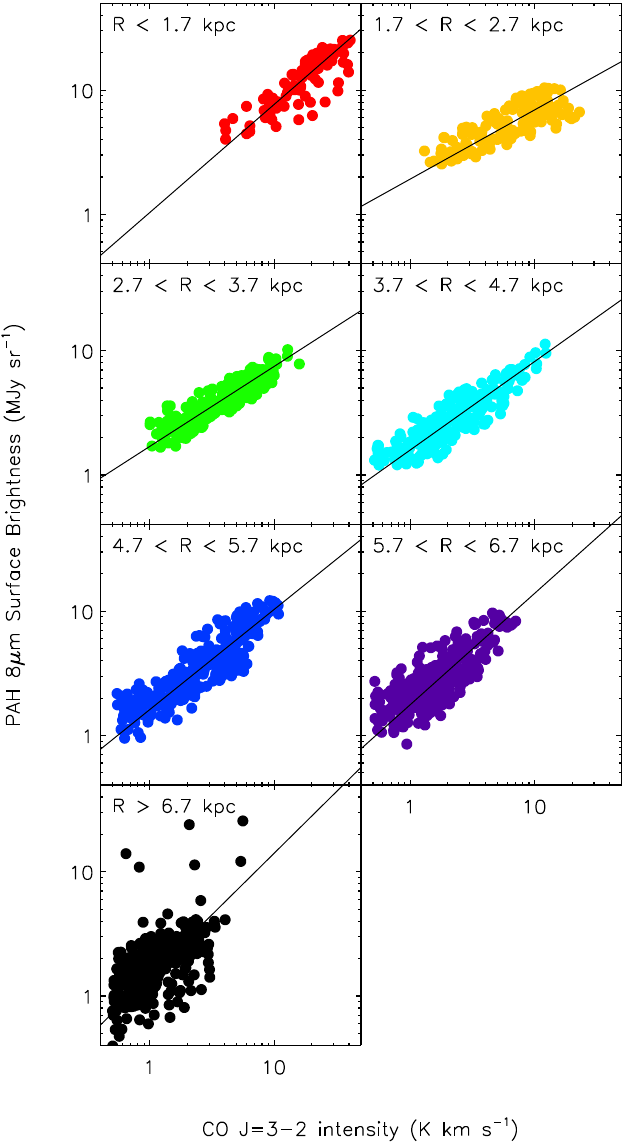}
\caption{Relationship between \cothree\/ intensity and \pah\/ surface brightness
at different galactocentric radii (over the entire galaxy disk where individual pixels are detected 
above 3$\sigma$). 
We colour-code the plotted points according to galactocentric radius as for Figure~\ref{fig:coratio-ico-fit}. 
The solid lines show bi-variate least squares fits, whose parameters are 
listed in Table~\ref{tab:fits}.
}
\label{fig:co32-pah-fit}
\end{figure}

In the top panel of Figure~\ref{fig:co32-pah-spiral} we plot the relationship between \cothree\/ and \pah\/ 
surface brightness, colour-coded for different radial intervals as in previous figures. 
We find a strong correlation, with a Pearson correlation coefficient of 
0.89 and the best-fitting bi-variate least squares relation is described by
\begin{equation}
\textrm{log} (\pah) = (0.718\pm0.009) \textrm{log} (\cothree) + (0.220\pm0.005)
\end{equation}. 
The largest scatter arises at the largest radii, both from the outermost parts of the 
galaxy disk and from inter-arm region II at large radii. 
The extreme outliers correspond to pixels in the outer-most regions of the disk.
The remainder of the scatter appears to arise from a more subtle dependence of the relationship 
on radius. We examine this further in Figure~\ref{fig:co32-pah-fit} by fitting separate 
linear relationships to different radial intervals, whose parameters are given in 
Table~\ref{tab:fits}. We find strong correlations (Pearson's correlation coefficients of 0.8--0.9) 
in every radial interval with the exception of the largest radii (0.45 at R$>$6.7~kpc). 
We note that the extreme outliers in the R$>$6.7~kpc radial bin correspond to pixels in the outer-most regions of the disk.

With the exception of the innermost bin, $R<1.7\,{\rm kpc}$, the slope of the relationship increases 
with increasing radius. The steep innermost bin corresponds to the region of the nuclear disk which is 
rich in molecular gas and exhibits high \cothree/\pah\/ ratios over most of the region, as seen in the left-hand panel 
of Figure~\ref{fig:maps-ratio}. 
The outliers below the fitted relation in this bin (red points) correspond to the region of highest \cothree/\pah\/ ratios 
(and also the region of highest CO arm-interarm contrast) described above.

The results in the present work support the idea that PAH emission is a good tracer of molecular gas, and that, for 
\mf\/ at least, the correlation between \cothree\/ and \pah\/ emission holds on sub-kpc scales 
and across different physical regions (whether in the nuclear region, spiral arms or inter-arm 
regions) in the galaxy. With the exception of one region of enhanced \cothree/\pah\/ ratio, 
the dependence of the relationship appears to be one of galactocentric radius.

One explanation for the relationship between \cothree\/ and \pah\/ emission could be if the 
formation of molecular clouds occurs in regions with stellar potential wells \citep{leroy08}. 
The CO emission then arises from the new molecular clouds and the increase in stellar emission 
in the potential wells heats the diffuse ISM, with enhanced PAH emission occurring in the regions 
where the strength of the interstellar radiation field is increased \citep[e.g.][]{bendo08}. If 
this is the case in \mf, then the strong correlation between PAH and CO emission on sub-kiloparsec  
scales suggests that the two processes are occurring in physically similar locations. 
Further evidence of this is the similar behaviour of \cothree\/ with 24\micron\/ emission, shown in 
the centre panels of Figures~\ref{fig:maps-ratio} and~\ref{fig:profiles-ratio}. The relationship between 
\cothree\/ and 24\micron\/ emission (lower panel of Figure~\ref{fig:co32-pah-spiral}, 
again colour coded for different radial intervals) is described by 
the best-fitting bi-variate least squares relation
\begin{equation}
\textrm{log} (24\micron) = (0.814\pm0.010) \textrm{log} (\cothree) + (0.257\pm0.006)
\end{equation}
with a Pearson correlation coefficient of 0.89. The relationship appears to vary with radius in a very similar way to 
that seen for \pah\/ emission. Again, the extreme outliers in the 
R$>$6.7~kpc radial bin correspond to pixels in the outer-most regions of the disk. 
An explanation of the decrease of both the \cothree/\pah\/ and 24\micron/\pah\/ ratios with radius 
would be if the PAHs are excited not only by UV photons (from massive star-forming regions and 
hence traced by \cothree\/ and 24\micron\/ emission) but also by softer optical photons which 
become relatively more important in the outer regions of the galaxy \citep[see e.g.][]{uchidaetal98,lianddraine02}.

\subsubsection{The relationship between \pah\/ and 24\micron\/ emission}

Previous authors have found that PAH emission does not trace star formation on sub-kiloparsec scales. 
We investigate this for the case of \mf\/ via the relationship between \pah\/ and 24\micron\/ 
surface density. 
It is thought that PAHs originate in the outflows of carbon-rich AGB stars and their descendants or in interstellar shocks 
\citep[e.g.][and references therein]{tielens08}. 24\micron\/ emission traces recent dust enshrouded
star formation, though may also include a component of diffuse emission that is not 
associated with recent star formation. We examine the relationship between \pah\/ and 
24\micron\/ emission in the right-hand panel of Figure~\ref{fig:maps-ratio} and the 
lower panel of Figure~\ref{fig:profiles-ratio}, 
which respectively show a map of the 
ratio of \pah\/ to 24\micron\/ surface density (using 
images that both have units of MJy~sr$^{-1}$) and the radial profile of this ratio.
The median ratio is 0.78, with a standard deviation of 0.28, and values for individual pixels 
range from $<$0.1 to 1.8.
There are several things that are immediately striking. 
First, the surface density of \pah\/ emission dominates over 24\micron\/ emission 
(i.e. \pah\/ to 24\micron\/ 
surface density ratios greater than unity) in the two outer spiral arms and in regions of 
prominent spurs bridging the two spiral arms. Elsewhere in the spiral arms there are are few 
pixels with values of about unity, but none within the central $\sim$2~kpc.
Second, apart from the two outer arms, regions dominant in \pah\/ emission appear to be 
concentrated to the south of the galaxy, away from the companion. 
Third, 24\micron\/ surface density is larger than the surface density of \pah\/ emission at 
the locations of all the bright 24\,\micron\/ peaks, but there are nonetheless some 24\micron\/ 
peaks at whose locations \pah\/ surface density appears to be dominate. 
It is also clear that the ratio of \pah\/ emission to 24\micron\/ emission increases with radius 
(lower panel of Figure~\ref{fig:profiles-ratio}), 
which is contrary to the findings of \citet{bendo10} for NGC\,2403.
Lastly, it is also striking that the variation of this ratio as a function of distance along the spiral arms and interarms 
is most similar to the radial profile of the \Hi\/ emission (see Figure~\ref{fig:spiral-profile}).

\section{Summary}
\label{sec:conclusion}
We have presented a new \cothree\/ molecular line map of the nearby grand-design spiral 
galaxy \mf\/ obtained with HARP-B on the JCMT. This map, covering an area of 
$\sim$14$\times$10\,kpc and having a spatial resolution of $\sim$600\,pc, is the first CO 
map in the \cothree\/ transition to cover the 
full extent of the disk in \mf. 

We have used the \cothree\/ map together with \cotwo\/ and \coone\/ data 
and other data from the literature to investigate the molecular gas properties of 
\mf\/ as a function of distance along the spiral arms, adopting a simple approach for 
defining the spiral structure.  We have found the following results:

\begin{itemize}
\item We detect \cothree\/ emission over an area of $\sim$21$\times$14\,kpc, 
across the entire disk of \mf\/ out to its companion NGC~5195.
The distribution of \cothree\/ emission follows a clear two-armed spiral structure with broad 
($\sim$1\,kpc) arms, and is in close agreement with the distribution of 
\cotwo\/ emission. We detect inter-arm \cothree\/ emission everywhere within a central area 
of $\sim$14$\times$10\,kpc, as well as several spurs, or bridges, of CO emission connecting 
the spiral arms.

\item We find that the contamination of submm continuum by in-band \cothree\/ emission 
ranges from up to 20\% in the spiral arms to $<$10\% in the inter-arm regions.

\item By projecting the CO integrated intensity maps of \mf\/ to various edge-on orientations, 
we have shown that the major axis profile of a galaxy with bright molecular gas rich spiral arms 
can mimic a molecular ring for some viewing angles.

\item For the \cothree\ and \cotwo\/ transitions there is a clear difference between the 
variation of arm and inter-arm emission with galactocentric radius, with the spiral arm 
emission decreasing with radius and the inter-arm emission relatively constant with radius. 
Conversely, for \coone\/ line and \Hi\/ emission the 
arm and inter-arm regions appears to follow a similar trend with radius.

\item The contrast between spiral arm and inter-arm \cothree\/ intensity  
decreases with galactocentric radius, falling by a factor of $\sim$3 over two kiloparsecs, 
There is a similar but less pronounced trend for \cotwo\/ line emission, but for \coone\/ and 
\Hi\/ emission the arm--inter-arm contrast appears relatively constant with radius.

\item At this spatial resolution, the CO~\jthree/\jtwo\/ line ratio varies from $\sim$0.1--1.0,  
with average values of 0.5 over the whole disk, 0.5 in the spiral arms, and 0.6 in the central 
$\sim$1\arcmin.  In the inter-arm regions, values range from 0.2--0.6.

\item There is no clear evidence that the CO line ratios vary as a function of distance along the 
spiral arms, but there does appear to be a trend for CO line ratios in the inter-arm regions to 
increase with radius.

\item We find a strong relationship between the \cothree\/ integrated intensity and the 
\pah\/ surface brightness. The ratio of \cothree\/ to \pah\/ emission decreases with radius, and 
we find a strong relationship between this ratio and the \cothree\/ intensity 
that varies with galactocentric radius.

\item A similar but slightly steeper relationship is found between \cothree\/ and 24\micron\/ surface 
brightness, with the ratio of \pah\/ to 24\micron\/ surface brightness increasing with radius.

\end{itemize}

\section*{Acknowledgments}
The James Clerk Maxwell Telescope is operated by The Joint Astronomy Centre on behalf of the Science and Technology Facilities Council of the United Kingdom, the Netherlands Organisation for Scientific Research, and the National Research Council of Canada. The HARP-B data was obtained under program ID M06BN005. This work made use of THINGS, `The HI Nearby Galaxy Survey' \citep{walter08}.

\bsp

\label{lastpage}

\end{document}